\documentclass[aps,prc,twocolumn,superscriptaddress,nofootinbib]{revtex4-1}

\usepackage{amssymb}
\usepackage{amsmath}
\usepackage{siunitx}
\usepackage{pifont}
\usepackage{multirow}
\usepackage{color}

\usepackage{physics}
\usepackage{graphicx}
\usepackage{dcolumn}
\usepackage{bm}
\usepackage{float}

  \newcommand {\nc} {\newcommand}
  \nc {\beq} {\begin{eqnarray}}
  \nc {\eeq} {\nonumber \end{eqnarray}}
  \nc {\eeqn}[1] {\label {#1} \end{eqnarray}}
  \nc {\eol} {\nonumber \\}
  \nc {\eoln}[1] {\label {#1} \\}
  \nc {\ve} [1] {\mbox{\boldmath $#1$}}
  \nc {\ves} [1] {\mbox{\boldmath ${\scriptstyle #1}$}}
  \nc {\mrm} [1] {\mathrm{#1}}
  \nc {\half} {\mbox{$\frac{1}{2}$}}
  \nc {\thal} {\mbox{$\frac{3}{2}$}}
  \nc {\fial} {\mbox{$\frac{5}{2}$}}
  \nc {\la} {\mbox{$\langle$}}
  \nc {\ra} {\mbox{$\rangle$}}
  \nc {\etal} {\emph{et al.}\ }
  \nc {\eq} [1] {(\ref{#1})}
  \nc {\Eq} [1] {Eq.~(\ref{#1})}
  \nc {\Sec} [1] {Sec.~\ref{#1}}
  \nc {\chap} [1] {Chapter~\ref{#1}}
  \nc {\anx} [1] {Appendix~\ref{#1}}
  \nc {\tbl} [1] {Table~\ref{#1}}
  \nc {\Fig} [1] {Fig.~\ref{#1}}
  \nc {\ex} [1] {$^{#1}$}
  \nc {\Sch} {Schr\"odinger }
  \nc {\flim} [2] {\mathop{\longrightarrow}\limits_{{#1}\rightarrow{#2}}}
  \nc {\IR} [1]{\textcolor{red}{#1}}
  \nc {\IB} [1]{\textcolor{blue}{#1}}
  \nc{\IG}[1]{\textcolor{green}{#1}}

\begin{document}


\title{Exploring core excitation in halo nuclei using halo effective field theory:\\
an application to the bound states of $^{11}$Be}

\author{Live-Palm Kubushishi}
 \email{lkubushi@ohio.edu}
\affiliation{%
Institut f\"ur Kernphysik, Johannes Gutenberg-Universit\"at Mainz,
55099, Mainz, Germany. 
}%
\altaffiliation[Current address: ]{Institute of Nuclear and Particle Physics and Department of Physics and Astronomy, Ohio University, Athens, OH 45701, USA%
}%
\author{Pierre Capel}%
 \email{pcapel@uni-mainz.de}
\affiliation{%
Institut f\"ur Kernphysik, Johannes Gutenberg-Universit\"at Mainz,
55099, Mainz, Germany. 
}%




\date{\today}

\begin{abstract}
Halo effective field theory (Halo-EFT) has proved to be very efficient for describing halo nuclei within models of nuclear reactions.
Its order-by-order expansion scheme enables us to single out the structure observables that are probed in reactions, and therefore improve the accuracy of their values inferred from experiment.
This formalism is however limited by its breakdown scale.
Structure effects beyond that scale cannot be considered explicitly in reaction models.
To extend the usual Halo-EFT, we include core excitation considering a particle-rotor model.
We apply it to the case of $^{11}$Be, the archetypical one-neutron halo nucleus. The corresponding set of coupled equations is solved using the R-matrix method on a Lagrange mesh. As a first application, we analyze in detail the structure of both bound states of $^{11}$Be and the $^{10}$Be-n phaseshifts at low energy in the corresponding partial wave as a function of the core deformation.
We also compute the electric dipole transition between these bound states.
The comparison of our results with existing \textit{ab initio} calculations, show that including core excitation within Halo-EFT can significantly improve the description of the $\half^+$ ground state of $^{11}$Be over single-particle models.
On the contrary, core excitation has a negative effect on the description of the $\half^-$ bound excited state. This is probably related to the presence of Pauli-forbidden states in our two-body model of $^{11}$Be.
\end{abstract}

\maketitle


\section{\label{sec:level1}Introduction}
The development of radioactive-ion beams (RIBs) in the mid-80s has enabled nuclear-physicists to explore the nuclear chart away from stability.
This technical breakthrough has led to the discovery of unexpected features among which the halo structure is the most exotic.
The main characteristic of halo nuclei is to exhibit a very large matter radius compared to their isobars \cite{Tan85b,Tan85l}.
This exceptional size is due to the very low separation energy these nuclei exhibit for one or two neutrons.
Thanks to this loose binding, these valence neutrons can tunnel far into the classically forbidden region, forming a sort of diffuse halo around a compact core, which contains most of the nucleons \cite{HJ87}.
The halo structure challenges usual nuclear models.
Accordingly, they are at the core of both experimental and theoretical studies \cite{Tan96}.

Halo nuclei are found close, or even at, the neutron dripline.
Being short lived, they cannot be studied through usual spectroscopic techniques, and most of the information on their structure is obtained through reactions measured at RIB facilities \cite{AN03}.
Knockout \cite{HT03}, transfer \cite{Wim18}, and breakup \cite{BC12} are the most used reactions to study halo nuclei.
To infer valuable nuclear-structure information from experimental cross sections, a reliable model of the reaction coupled to a realistic, yet numerically tractable, description of the projectile is needed.
Most reaction models are based on a few-body approach, where the halo nucleus is described as a two- or three-body quantal system, and the interaction of the nucleus components with the target is simulated by optical potentials \cite{AN03,BC12}.

Halo nuclei exhibit a clear separation of scales with one or two valence neutrons loosely bound to a tightly bound core.
This characteristic implies that they can be well described within an effective-field theory (EFT): Halo-EFT \cite{bertulani2002review}; see Ref.~\cite{Hammer17review} for a recent review.
In this EFT, the Hamiltonian that describes the core-halo structure is expanded over the small parameter $\eta$ obtained by dividing the small radius of the former $R_c$ by the broad size of the latter.
For $^{11}$Be, the archetypical one-neutron halo nucleus, $\eta\approx0.4$ \cite{HP11}, hence small enough for a quick convergence of the EFT expansion.
At each order of the expansion, a new term is added to the effective core-neutron potential, providing finer and finer details to its description.

At leading order (LO), the core-neutron interaction reduces to a simple contact term in the partial wave that hosts the nucleus' bound state.
For most halo nuclei, and in particular for $^{11}$Be, that partial wave is the $s$ wave.
The depth of that term is usually fitted to reproduce the one-neutron separation energy of the nucleus.

At next-to-leading order (NLO), a new term is added to that contact force and in other partial waves, typically in the $p$ waves.
The magnitude of these additional terms are adjusted to reproduce other observables of the nucleus, such as the asymptotic normalization coefficient (ANC) for bound states or the first terms of the effective-range expansion in the continuum.
The values of these observables are taken from experiment or from accurate theoretical predictions.
In particular for $^{11}$Be, an \emph{ab initio} calculation has been performed by Calci \etal within the no-core shell model with continuum (NCSMC) \cite{Calcietal16}.

The Halo-EFT expansion scheme not only provides an effective few-body description of halo nuclei, but it also ranks systematically the structure observables, upon which it is fitted, in order of significance.
This is very valuable in nuclear-reaction theory, because it provides a reliable estimate of the relative influence of these observables upon reaction cross sections.
For example, in breakup calculations, the binding energy is the most important parameter, followed by the ANC of the bound state, and the phaseshift in the low-energy continuum \cite{Capeletal18,MC19,MYC19,CP23}.
Accordingly, Halo-EFT provides a systematic framework within which results obtained in past analyses fit nicely, see, e.g., Refs.~\cite{CN06,SN08,Esb09}.
Similar results have also been obtained for transfer \cite{YC18,MYC19} and knockout \cite{HC19,HC21}.
Coupled to accurate models of reaction, Halo-EFT provides a reliable way to infer structure observables from reaction measurements.
When the low-energy constants (LECs) of this description are fitted to \emph{ab initio} calculations, Halo-EFT enables us to bridge directly these theoretical predictions to experimental reaction data.

In some cases, this expansion scheme is not sufficient to capture the whole nuclear-structure physics that affects reaction observables.
This is especially true for effects that lie beyond the breakdown scale of the EFT.
For example, the cross section for the breakup of $^{11}$Be on $^{12}$C cannot be fully described considering the sole single-particle description of $^{11}$Be within Halo-EFT \cite{Capeletal18,MC19}.
Even at next-to-next-to-leading order (N$^2$LO), there is a clear missing strength in the vicinity of the $\fial^+$ and $\thal^+$ resonances of $^{11}$Be \cite{KC25}.
In Ref.~\cite{ML12}, Moro and Lay have shown that to properly reproduce the experimental cross section at these resonance energies, the internal structure of the core can no longer be ignored.
In particular, the excitation of the $^{10}$Be core to its first $2^+$ state seems key to explain the reaction.
This was confirmed within an EFT approach using a three-body force to describe the virtual excitation of the core in the reaction \cite{CPH22}.

Because of the aforementioned advantages, it is useful to include this core degree of freedom within Halo-EFT.
This will enable us to treat explicitly the core excitation within a few-body model of the reaction.
Moreover, it will extend the range of validity of Halo-EFT by shifting its breakdown scale to higher energy.
We reach this goal in the present paper.
Similar to previous approaches \cite{VinhMau95,Esbensen1995,Nunes1996,ThompsonFaCE}, we consider a collective model of the core to which one neutron is loosely bound.
In particular, we base our developments upon the particle-rotor model of Bohr and Mottelson \cite{bohrmottelson}.
In that vision, the first states of the core are seen as members of the same rotational band.
This enables us to include the low-lying states of the core within the description of the halo nucleus to effectively extend the usual single-particle vision of halo nuclei.
This effective approach to study light deformed nuclei is to some extent in continuation of previous studies of deformation effects in heavy odd-mass nuclei from an EFT perspective \cite{PW20,ACP22}.
Along these lines, Coello Pérez and Papenbrock recently published an extensive review \cite{CPP25} on effective field theories for collective excitations of atomic nuclei summarising the building blocks of these approaches and comparing them with current nuclear models.
The aim of our approach is to expand the panorama outlined in this review to light deformed halo nuclei.

Once developed and implemented, this rotor-EFT is used on the particular case of $^{11}$Be.
Besides providing a simple and elegant coupled-channel description of $^{11}$Be, this particle-rotor vision is also justified on physical grounds.
Calculations within the antisymmetrized molecular dynamics (AMD) of $^{10}$Be indicates that its $0_1^+$, $2_1^+$, and $4_1^+$ states belong to the same rotational band \cite{KHD99}.
This has been confirmed in a recent no-core shell model (NCSM) calculation of the neutron-rich isotopes of beryllium by McCoy \cite{McCoy24}.

As in Ref. \cite{Capeletal18}, we fit the LECs of that description to the experimental one-neutron separation energy and the ANC predicted by Calci \etal\ \cite{Calcietal16}.
For each bound state, we study how core excitation influences our calculations and if it improves the few-body description of $^{11}$Be compared to the many-body model of Calci \etal
Interestingly, this provides a way to analyze in rather simple terms the results of these \emph{ab initio} calculations leading to a better understanding of the structure of $^{11}$Be.

This article is structured as follows.
In \Sec{model}, we describe the particle-rotor model within Halo-EFT.
Its application to the case of $^{11}$Be is then detailed in \Sec{results}.
We analyze specifically the description of both bound states: the $\half^+$ ground state in \Sec{gs}, and the $\half^-$ excited state in \Sec{es}.
The strong electric dipole transition between these two states is computed within that model in \Sec{E1}.
We conclude our analysis in \Sec{conclusion}. In the Appendices \ref{type2} and \ref{s/=sc}, we explore the model state further.
This exploration has enabled us to find other sets of solutions, which are of no physical significance, but which we present for completeness.

\section{\label{model}Halo-EFT particle-rotor model}

Within the usual Halo-EFT, one-neutron halo nuclei are described as two-body quantal structures with a halo neutron $n$ loosely bound to a core $c$ \cite{bertulani2002review,Hammer17review}.
Truth to the EFT approach, the short-range physics is ignored, and the $c$-$n$ interaction is simulated by a contact potential and its derivatives.
In particular the internal structure of the core is neglected.
To include core excitation, we follow previous approaches \cite{VinhMau95,Esbensen1995,Nunes1996, ThompsonFaCE} and describe the core within a collective model of nuclei \cite{bohrmottelson}.
Assuming that the core suffers a permanent deformation, its spectrum can be described either as a vibrator \cite{VinhMau95} or as a rigid rotor \cite{Esbensen1995,Nunes1996, ThompsonFaCE}.
This leads to the following core-neutron Hamiltonian
\begin{equation}
 H(\ve{r},\xi)= -\frac{\hbar^2}{2\mu}\Delta + V_{cn}({\ve{r}},\xi) + h_{c}({\xi}),
 \label{H_tot}
\end{equation}
where \ve{r} is the core-neutron relative coordinate, $\mu = m_cm_n/(m_c+m_n)$ is the $c$-$n$ reduced mass, with $m_c$ and $m_n$ the masses of the core and the valence neutron, respectively.
The $c$-$n$ interaction is simulated by an effective potential $V_{cn}$, which, through deformation, depends on $\xi$, the internal coordinates of the core.
The core structure is accounted for by the Hamiltonian $h_c$, which has the following eigenstates of discrete eigenenergies $\epsilon_c^{I_c^{\pi_c}}$
\begin{equation}
   h_{c}({\xi})\phi_{M_{c}}^{I_{c}^{\pi_c}}(\xi)=\epsilon_{c}^{I_c^{\pi_c}} \phi_{M_{c}}^{I_{c}^{\pi_c}}(\xi)
   \label{H_core}
\end{equation}
where $\pi_c$ is the parity of the core, $I_c$ is its spin and $M_c$ is the spin projection.
In the present approach, we consider a rotational model of the core \cite{bohrmottelson,Esbensen1995,Nunes1996, ThompsonFaCE}.
In that model, the core is seen as an axially symmetric rigid rotator.
Its internal coordinates $\xi$ are thus the Euler angles, and the core wave functions $ \phi_{M_{c}}^{I_{c}\pi_c}$ are expressed as a function of the Wigner matrices ${\cal D}^{I_c}_{M_cK_c}$ \cite{bohrmottelson}
\beq
\phi_{M_{c}}^{I_{c}\pi_c}({\xi})&=&\sqrt{\frac{2I_c+1}{16\pi^2}}i^{\frac{1-\pi_c}{2}}\nonumber\\
 &\times&\left[{\cal D}^{I_c}_{M_cK_c}({\xi})+\pi_c(-1)^{I_c+K_c}{\cal D}^{I_c}_{M_c\,-K_c}({\xi})\right],
\eeqn{e2a}
with $K_c$ the projection of the spin in the intrinsic rest frame of the core.
It is a constant value within a rotational band, so we ignore it in the notation of the core eigenstates \eq{H_core}.
The eigenstates $\Psi^{J^\pi M}$ of the Hamiltonian \eqref{H_tot} are defined by their total angular momentum $J$ and their parity $\pi$, the last quantum number $M$ is the projection of $J$.
They can be expanded in a basis separating the internal structure of the core from the $c$-$n$ relative motion
\begin{equation}
    \Psi^{J^\pi M}(\ve{r},\xi)=\sum_{\alpha} i^\ell \frac{u_{\alpha}(r)}{r} [\mathcal{Y}_{\ell j}(\hat{r}) \otimes \phi_{M_{c}}^{I_{c}\pi_c}(\xi)]^{J M},
    \label{totalwf}
\end{equation}
where $\alpha$=$\left\{n_r,\ell,s,j,I_c,\pi_c\right\}$ defines a channel, i.e., a set of quantum numbers describing each allowed core-neutron configuration.
In details, $n_r$ is the number of nodes in the reduced radial wave function $u_\alpha$, $\ell$ is the orbital angular momentum of the $c$-$n$ relative motion, $s=\frac{1}{2}$ is the neutron spin, and $j$ is the angular momentum obtained from the coupling of $\ell$ and $s$: $\ve{j} = \ve{\ell} + \ve{s}$.
This angular momentum $j$ is then coupled to the spin of the core $I_{c}$ to obtain the total angular momentum $J$: $\ve{J} = \ve{j} + \ve{I_{c}}$.

In \Eq{totalwf}, 
$\mathcal{Y}_{\ell jm}$ describes the spin-angular part of the channel wave function: $\mathcal{Y}_{\ell jm}=[ Y_{\ell} \otimes \chi_s]^{j m}$, which couples the spherical harmonics $Y_{\ell}^{m_\ell}$ to the spinors $\chi_s^{m_s}$.
The magnitude of each configuration $\alpha$ in the total wave function is evaluated by its spectroscopic factor (SF)
\begin{equation}  {\cal S}_{\alpha}=\int_0^{\infty}\left|u{_\alpha}(r)\right|^2 dr.
\label{defSF}
\end{equation}
Since the wave functions $\Psi^{J M}$ are normed to unity, the SFs of all the channels add up to one: $\sum_{\alpha} {\cal S}_{\alpha}=1$.
By plugging the expansion \eqref{totalwf} in the Schrödinger equation for the $c$-$n$ system, we obtain a set of coupled-channel equations for the reduced radial wavefunctions
\begin{eqnarray}
\left\{\frac{\hbar^2}{2\mu}\left[-\frac{d^2}{dr^2}+\frac{\ell(\ell+1)}{r^2}\right]+V_{\alpha \alpha}(r) + \epsilon_{\alpha}- E\right\} u_{\alpha}(r) =\nonumber \\
 -\sum_{\alpha'\neq \alpha}V_{\alpha \alpha'}(r)u_{\alpha'}(r), \hspace{13mm}
    \label{coupled_eq}
\end{eqnarray}
where the matrix elements of the $c$-$n$ interaction read
\beq
V_{\alpha \alpha'}(r) = \mel{\mathcal{Y}_{\alpha}(\hat{r})  \phi_{\alpha}(\xi)}{V_{cn}(\ve{r},\xi)}{\mathcal{Y}_{\alpha'}(\hat{r}) \phi_{\alpha'}(\xi)}.
\eeqn{e6}

The non-diagonal matrix elements govern the coupling between the different channels.
This coupling is induced by the deformation of the core,
which translates into an angle dependence of the $c$-$n$ interaction $V_{cn}$.
Within the Halo-EFT spirit, we treat this deformation perturbatively to obtain the following expression
\begin{equation}
    V_{cn}(\ve{r},\xi)= V(r;\sigma) + \beta\, \sigma\, Y_{2}^{0}(\hat{r}') \frac{d}{d\sigma}V(r;\sigma)
    \label{Vcpl_order1}
\end{equation}
where $\hat{r}'$ is the solid angle of the $c$-$n$ relative coordinate $\ve{r}'$ expressed in the core's intrinsic reference frame, $\beta$ is the deformation parameter, and $\sigma$ is the range of the potential $V$, corresponding to the undeformed $c$-$n$ interaction.

In previous work, $V$ was typically of a Woods-Saxon form \cite{Esbensen1995,Nunes1996, ThompsonFaCE}.
Within the present approach, true to Halo-EFT, we consider contact terms plus their derivatives, which we regularize with Gaussians.
In particular, we choose the NLO interaction considered in Ref.~\cite{Capeletal18}
\begin{equation}
    V(r;\sigma)=V_{J^\pi}^{(0)} e^{-\frac{r^2}{2\sigma^2}} + V_{J^\pi}^{(2)} r^2  e^{-\frac{r^2}{2\sigma^2}}.
    \label{Vcn_NLO_central}
\end{equation}
The width of these Gaussians $\sigma$ is an unfitted parameter which corresponds to the short-range physics neglected in this EFT approach.

The interaction \eq{Vcn_NLO_central} contains two low-energy constants (LECs), $V_{J^\pi}^{(0)}$ and $V_{J^\pi}^{(2)}$, which are adjusted for each value of the total angular momentum $J$ and parity $\pi$ to reproduce know structure observables of the nucleus.
Accordingly, they are identical in all the channels of a given state \eq{totalwf}.
For the bound states, which are the focus of this paper, we consider as a first constraint the one-neutron separation energy $S_n(J^\pi)$, which is known experimentally \cite{Ajzen1990}.
As a second constraint, we use the ANC predicted by Calci \etal \cite{Calcietal16}.
For continuum states, these LECs can be fitted to the $c$-$n$ phaseshift predicted by this same \emph{ab initio} calculation.
This is the subject of a future paper \cite{KC25c}.

A simple and efficient method to solve the set of coupled-channel equations \eqref{coupled_eq} is the R-matrix method on a Lagrange mesh \cite{Descouvemont2010rmat,Baye2015lagrange}.
This method is both simple and robust to treat bound and scattering states on the same footing \cite{hesseJMS1998}.
We have developed a dedicated computer code to solve that set of coupled equations numerically.
This allows us in particular to calculate the radial wavefunctions $u_{\alpha}$ and their respective ANC for the bound states as well as the collision matrix $U^{J^{\pi}}$ in the corresponding continua.

\begin{figure*}[ht] 
    \centering
    \begin{minipage}{0.50\linewidth}
        \centering
        \includegraphics[width=\linewidth]{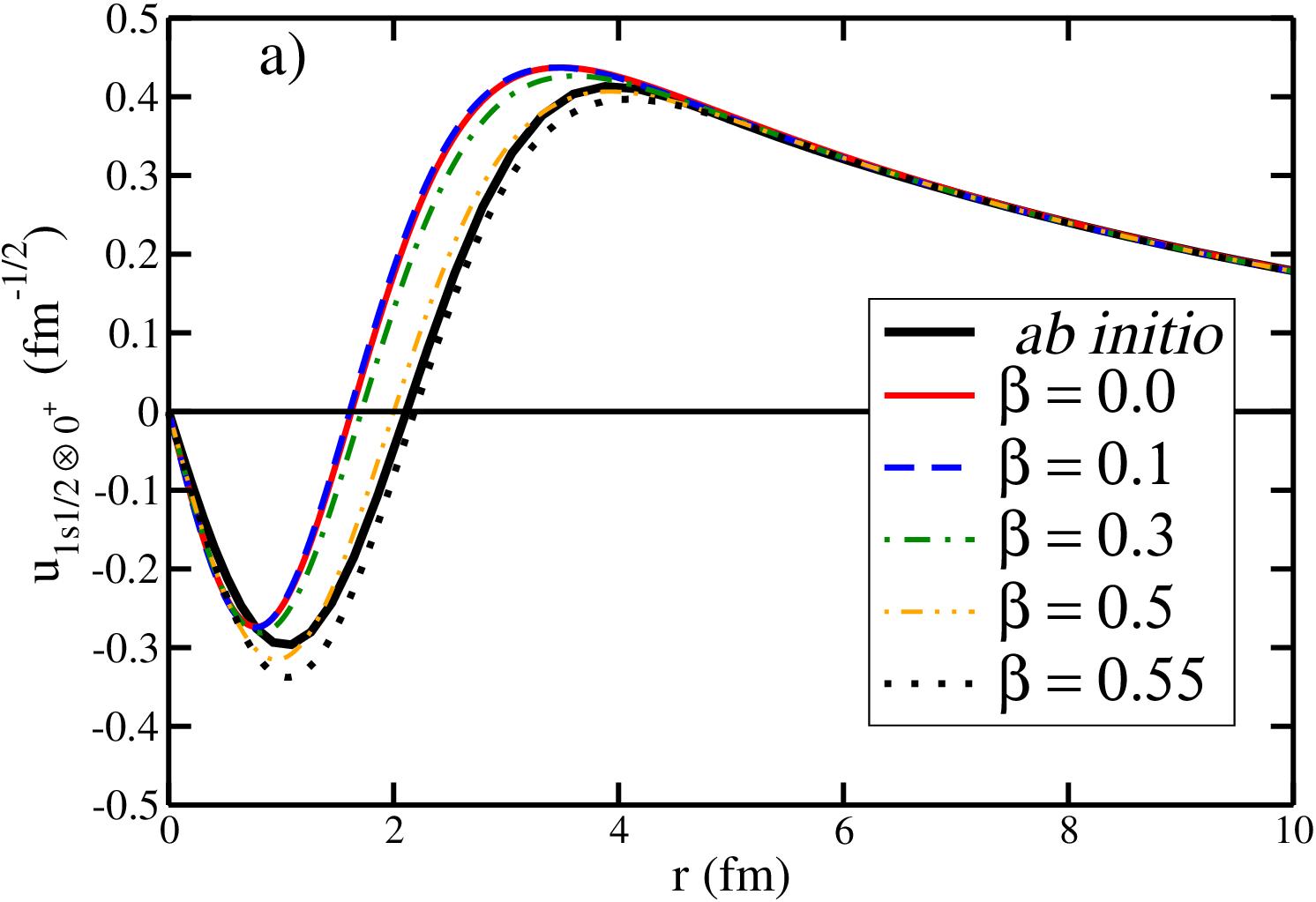}
    \end{minipage}\hfill
    \begin{minipage}{0.50\linewidth}
        \centering
        \includegraphics[width=\linewidth]{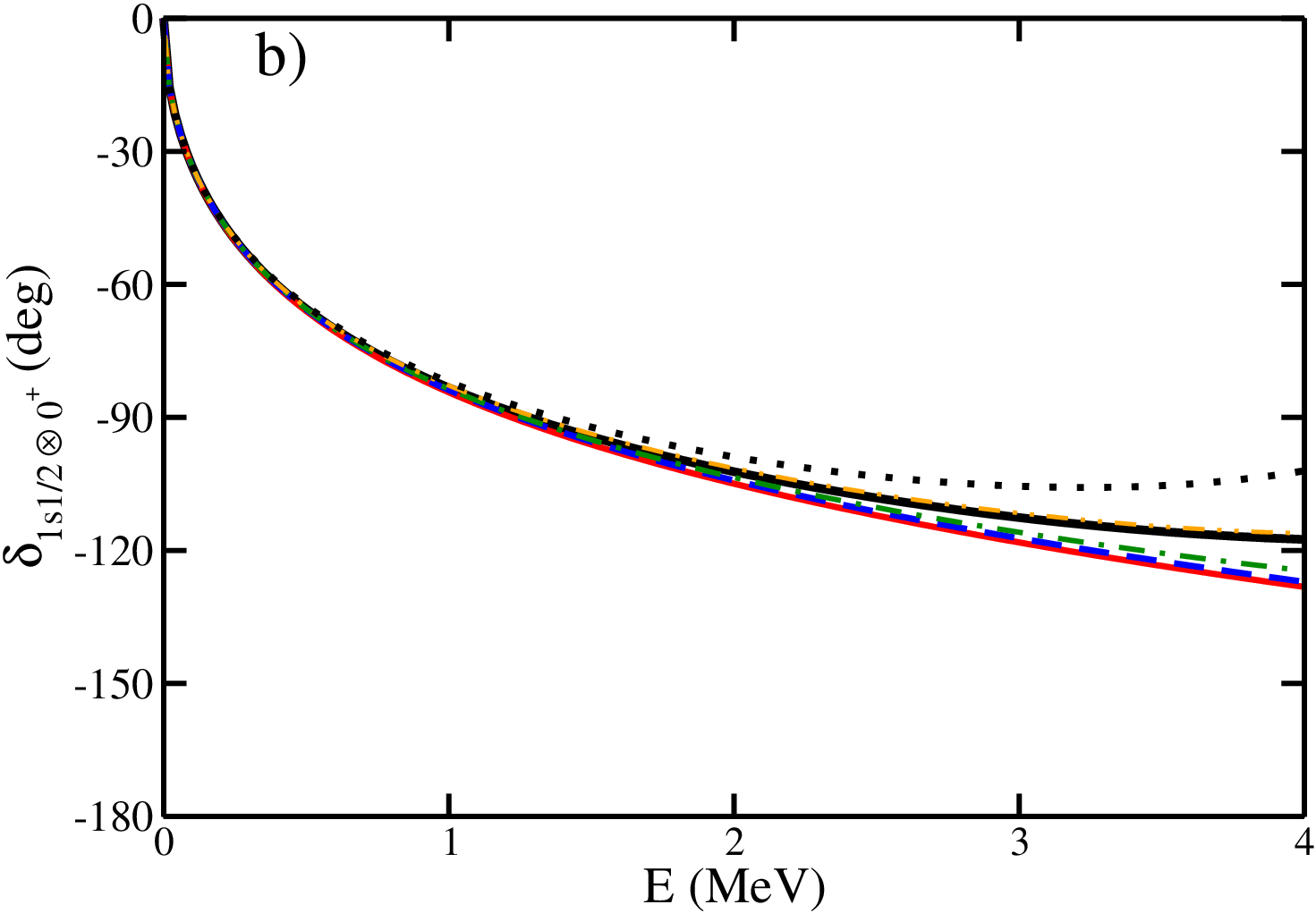}
    \end{minipage}
    \caption{Halo-EFT particle-rotor calculation of the $\frac{1}{2}^+$ ground state of $^{11}$Be for different values of the deformation parameter $\beta$ using $\sigma=1.5$~fm. (a) Radial wave functions in the $1s_{1/2} \otimes 0^+$ channel. (b) Corresponding eigenphaseshifts as a function of the $^{10}$Be-$n$ relative energy $E$.
    The \textit{ab initio} predictions are plotted in thick black lines \cite{Calcietal16}.}
    \label{fig:gs_wf_up_15_15}
\end{figure*}

From these coupled-channels calculations, we study the impact of core excitation on both the bound-state wavefunctions and the $c$-$n$ scattering phaseshifts in the corresponding partial wave, and compare them to the \emph{ab initio} calculations of Ref.~\cite{Calcietal16}.
This is what we present in \Sec{results}.
For each bound state of $^{11}$Be, we adjust the LECs of the $c$-$n$ interaction for different values of the deformation $\beta$, and we confront the radial wave functions and the corresponding phase shift in the continuum with the prediction of Calci \etal \cite{Calcietal16}.

Similar to previous work \cite{Esbensen1995,Nunes1996}, we consider only the first two states of $^{10}$Be: its $0^+$ ground state and its first $2^+$ excited state at $\epsilon_{2^+}=3.368$~MeV \cite{Ajzen1990}.
We assume these states to be members of a $K_c^{\pi_c}=0^+$ rotational band; a hypothesis confirmed by AMD \cite{KHD99} and NCSM \cite{McCoy24} calculations.

\section{\label{results} Results and analysis}

\subsection{Ground state: 1/2$^+$}\label{gs}
$^{11}$Be presents a $\frac{1}{2}^+$ ground state  with an energy of $E_{1/2^+}=-0.503$~MeV\footnote{The most recent value for that one-neutron separation energy is slightly different: $S_n(^{11}{\rm Be})=0.50164$~MeV$\pm0.25$~keV \cite{Masses21}. Nevertheless, we consider the same value as in \cite{Capeletal18} to ease the comparison with these previous Halo-EFT calculations.} \cite{Ajzen1990} and an ANC ${\cal C}_{1/2^+}=0.786$~fm$^{-1/2}$ predicted from the \textit{ab initio} calculation of Calci \etal \cite{Calcietal16}.
That particular value of the ANC has been selected first because it is reliable: it leads to excellent agreement with various peripheral reactions involving $^{11}$Be: breakup \cite{Capeletal18,MC19}, transfer \cite{YC18,deltuva23}, and knockout \cite{HC21}.
Second, we will compare our results to that many-body calculation to test how core excitation can improve the few-body model of halo nuclei.
We chose to reproduce both observables to the nearest three significant digits.

After coupling all angular momenta defined by Eq.~\eqref{totalwf}, the $\frac{1}{2}^+$ ground state of $^{11}$Be can be written as:
\begin{equation}
\ket{\frac{1}{2}^+}=\ket{1s_{1/2} \otimes 0^+} + \ket{0d_{5/2} \otimes 2^+} + \ket{0d_{3/2} \otimes 2^+}.
\label{overlap_wf_gs}
\end{equation}

To study the effects of the coupling between these channels, we perform calculations within the model described in Sec.~\ref{model} with $\sigma=1.3$, 1.5, 1.8, and 2.0~fm.
This range is limited on its lower end by the Wigner bound \cite{Wigner_wignerbound,PHILLIPS1997,HAMMER2010}; below that value it is not possible to fit both the binding energy and the ANC for all values of the deformation $\beta$.
The upper bound is limited by the radius of the $^{10}$Be core $R_c\sim2.2$--2.4~fm \cite{Tan85l,descouvemont97,Arai2010}.

\begin{figure*}[ht]
    \centering
    \begin{minipage}{0.50\linewidth}
        \centering
        \includegraphics[width=\linewidth]{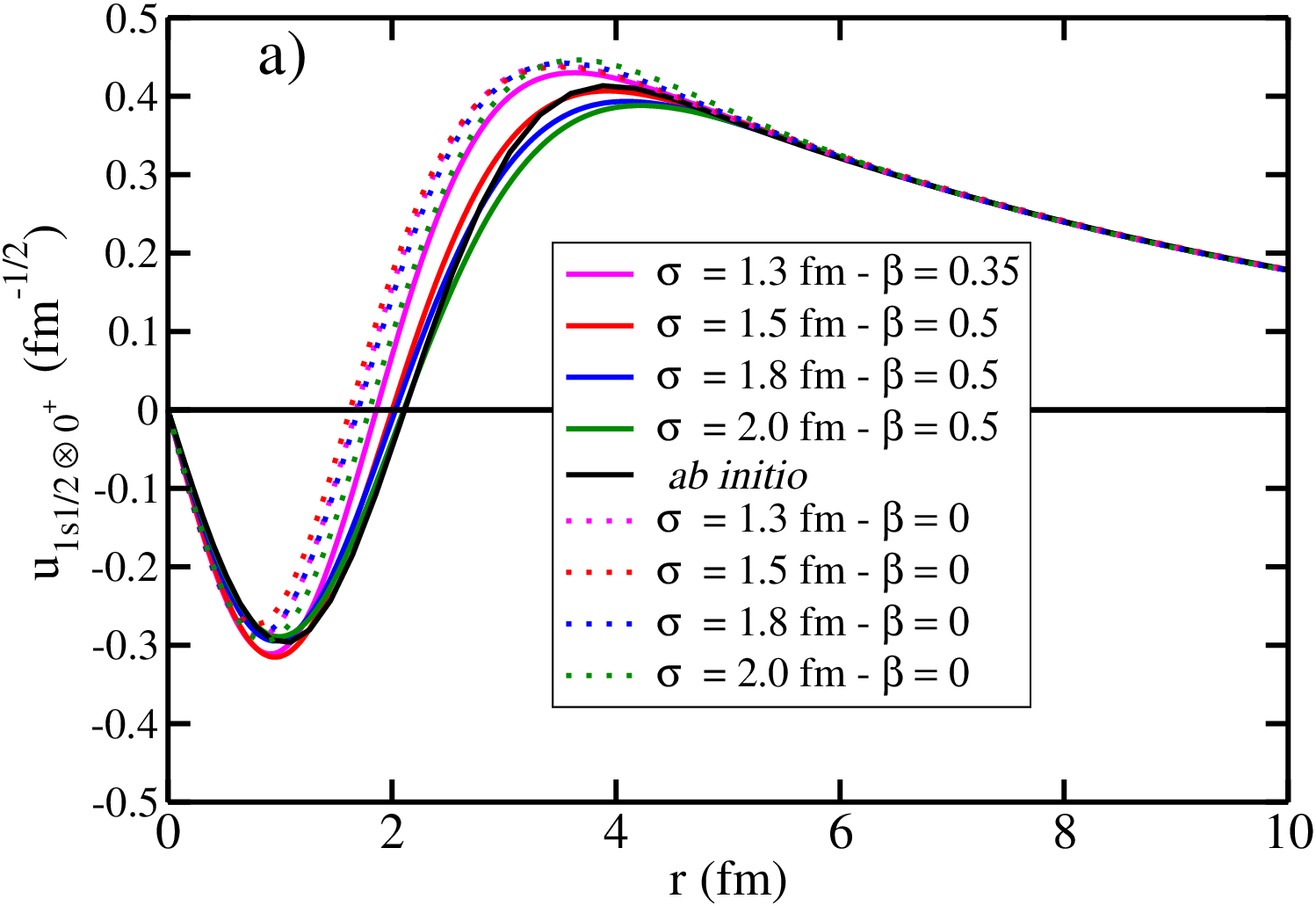}
    \end{minipage}\hfill
    \begin{minipage}{0.50\linewidth}
        \centering
        \includegraphics[width=\linewidth]{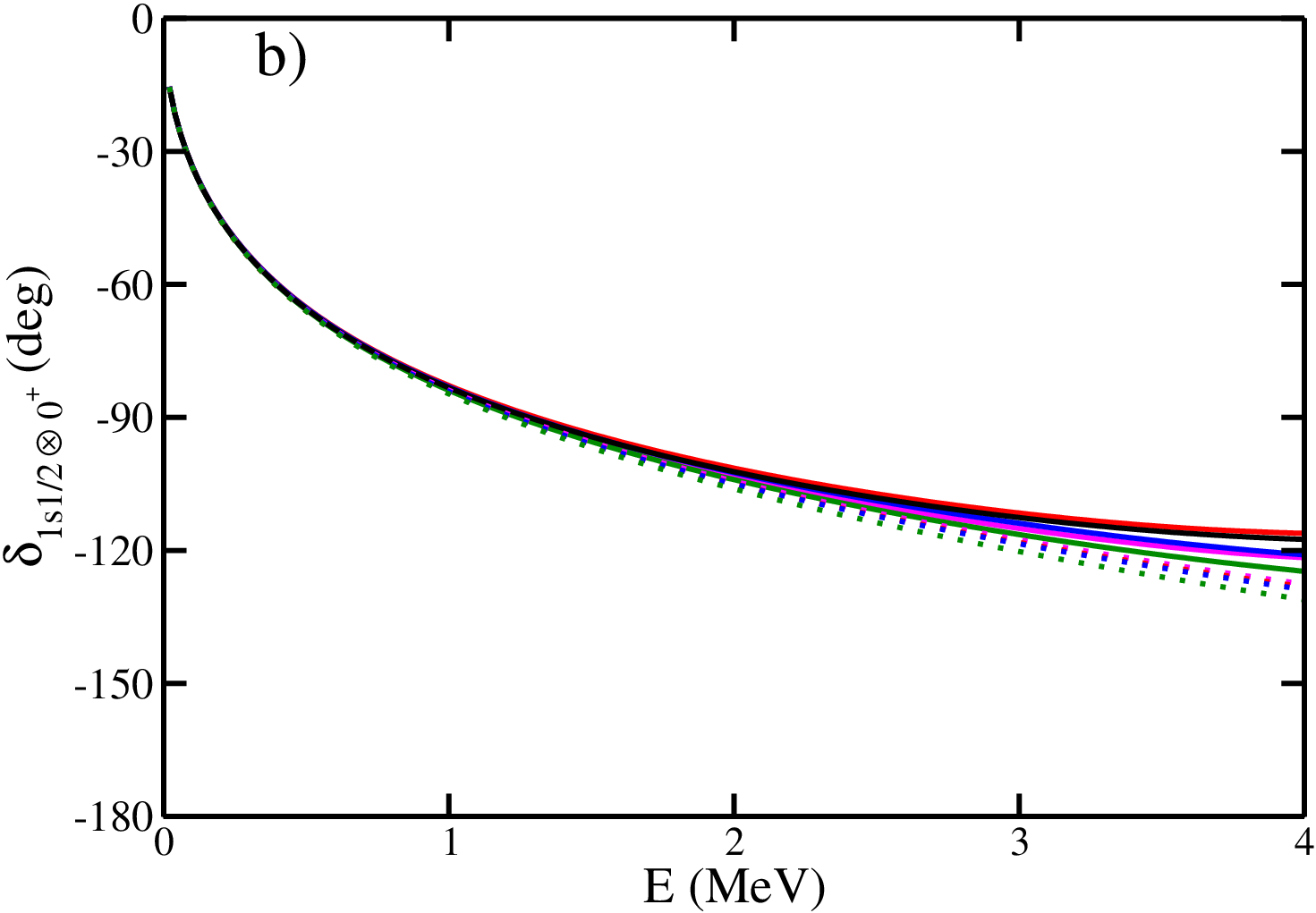}
    \end{minipage}
    \caption{(In)sensitivity of our calculations of the $\frac{1}{2}^+$ ground state of $^{11}$Be to the cutoff $\sigma$.
    The solid lines correspond to the solutions with the coupling $\beta \sim 0.5$ that best fits the \emph{ab initio} prediction (thick black line).
    The dotted lines correspond to Halo-EFT solutions without coupling ($\beta=0$).
        (a) Radial wave function in the dominant $1s_{1/2} \otimes 0^+$ channel.
        (b) Eigenphaseshifts in the $s_{1/2} \otimes 0^+$ channel as a function of the $^{10}$Be-$n$ relative energy $E$.}
    \label{fig:best1/2+_type1solution}
\end{figure*}

Figure~\ref{fig:gs_wf_up_15_15} presents the results obtained for $\sigma = 1.5$~fm; similar results are obtained for other values of the potential range.
Panel (a) shows the reduced radial wave function of $^{11}$Be's ground state in its dominant $1s_{1/2} \otimes 0^+$ channel as a function of $r$ for different values of the deformation parameter $\beta$.
The corresponding phaseshifts are displayed in panel (b) as a function of the $^{10}$Be-$n$ relative energy $E$.
The parameters of the effective $^{10}$Be-$n$ potential \eqref{Vcn_NLO_central} are listed in Table~\ref{tab:gs_wf_up_15_15} as a function of $\beta$, alongside the structure observables they produce.
The \textit{ab initio} predictions of \cite{Calcietal16} are plotted for comparison with thick black lines in both panels of Fig.~\ref{fig:gs_wf_up_15_15} and listed at the bottom of Table~\ref{tab:gs_wf_up_15_15}.

\begin{table}[h]
\caption{\label{tab:gs_wf_up_15_15}
Parameters of the effective $^{10}$Be-$n$ potential \eqref{Vcn_NLO_central} in the $\frac{1}{2}^+$ ground state for $\sigma=1.5$~fm.
For each deformation parameter $\beta$, we provide the depths $V^{(0)}_{1/2^+}$ and $V^{(2)}_{1/2^+}$, which reproduce the listed $E_{1/2^+}$ energy, ANC, and SF.
The \textit{ab initio} predictions of Ref.~\cite{Calcietal16} are given in the last line.}

\begin{ruledtabular}
\begin{tabular}{cccccccc}
 $\beta$ & $V^{(0)}_{1/2^+}$ & $V^{(2)}_{1/2^+}$ & $E_{1/2^+}$ & ${\cal C}_{1/2^+}$ & ${\cal S}_{1s_{1/2} \otimes 0^+}$ \\ 
 & (MeV) & (MeV fm$^{-2}$) & (MeV) & (fm$^{-1/2}$) & \\ \hline
0  & $-103.610$ & 0.8 & $-0.5031$ & 0.7857   & 1.00  \\
0.1 & $-102.309$ & 0.6 & $-0.5031$ & 0.7858  & 0.99 \\
0.3 & $-91.127$  & $-1.2$ & $-0.5031$ & 0.7864  & 0.97 \\
0.5 & $-66.144$  & $-6.3$ & $-0.5031$ & 0.7862 & 0.92 \\
0.55 & $-51.874$  & $-10.3$ & $-0.5031$ & 0.7861 & 0.90  \\  \hline
\textit{ab initio} & \cite{Calcietal16}  &   & $-0.5$ & 0.786 & 0.90  \\ 
\end{tabular}
\end{ruledtabular}
\end{table}

The results displayed in \Fig{fig:gs_wf_up_15_15}(a) show that increasing the coupling strength $\beta$ leads to significant changes in the interior of the wave function but leaves the asymptotics unchanged since, by construction, all the wave functions have been fitted to the same binding energy and ANC.
Obviously the SF ${\cal S}_{1s_{1/2}\otimes0^+}$ decreases with $\beta$ because part of the probability strength moves to the other channels (see Table~\ref{tab:gs_wf_up_15_15}). 
In addition, the node and maximum of the radial wave function move to larger radii.
At $\beta=0.5$ (orange dash-dot-dotted line), our rotor-EFT calculation matches nearly perfectly the \emph{ab initio} prediction.
Accordingly, the SF obtained for this $\beta$ is very close to the \emph{ab initio} one.
Beyond that $\beta$, the rotor-EFT overlap wave function moves further to the right and the SF decreases below 0.9. Note that beyond $\beta=0.55$, it is difficult to find a solution that reproduces both the energy and ANC of the bound state.
This is probably due to the existence of a second solution, which we discuss in \anx{type2}.

This interesting result is also observed in the $s_{1/2}\otimes 0^+$ eigenphaseshift displayed in \Fig{fig:gs_wf_up_15_15}(b).
Already in the single-channel Halo-EFT of Ref.~\cite{Capeletal18}, the behavior of the \emph{ab initio} phaseshift is well reproduced up to $E=1.5$~MeV, see the $\beta=0$ red solid line.
This is ensured because the $c$-$n$ interaction is fitted to reproduce the binding energy and the ANC predicted by Calci \etal \cite{JMScapel2010}.
Interestingly, this property seems to be conserved within our coupled-channel approach, even though this does not seem to always be the case, see \anx{type2}.
Moreover, introducing the coupling to core excitation leads to an increase of $\delta_{s1/2\otimes 0^+}$ at larger energies, which improves the agreement with the \emph{ab initio} calculations.
Here too, at $\beta=0.5$, our result matches perfectly those of Calci \etal up to $E=4$~MeV.
Beyond that $\beta$, the rotor-EFT phaseshift overestimates the \emph{ab initio} one.

These results can be qualitatively understood from the changes observed in the potential depths; see Table~\ref{tab:gs_wf_up_15_15}.
Within the particle-rotor model, the effect of coupling is globally attractive \cite{CDN10}.
To compensate it and reproduce the correct binding energy, the central depth $V^{(0)}_{1/2^+}$ has to be reduced when $\beta$ increases.
To then obtain the desired ANC, the surface term of the potential \eqref{Vcn_NLO_central} needs to be more attractive.
The combined effect of the reduction of the central term, the increase of the surface term, and the additional attraction due to the coupling leads to the observed shift of the wave function to larger radii.
The additional attraction at large $r$ also increases $\delta_{s1/2\otimes 0^+}$, so that it moves towards the \emph{ab initio} results.
The results presented above are nearly independent of $\sigma$ and $\beta$.
To illustrate this, we display in Fig.~\ref{fig:best1/2+_type1solution} the optimal solutions obtained with cutoffs $\sigma=1.3$, 1.5, 1.8, and 2.0~fm.
The solid lines show these solutions at the value of $\beta$ that best fits the \emph{ab initio} predictions; see Table~\ref{tab:gs_optimal_cases}.
The dotted lines show the results without coupling, i.e., within the usual single-channel Halo-EFT of \cite{Capeletal18}.

\begin{table}[b]
\caption{\label{tab:gs_optimal_cases}
Parameters of the effective $^{10}$Be-$n$ potential \eqref{Vcn_NLO_central} in the $\frac{1}{2}^+$ ground state.
For each cutoff $\sigma$, we provide the depths $V^{(0)}_{1/2^+}$ and $V^{(2)}_{1/2^+}$, at the value of $\beta$, which best reproduces the \textit{ab initio} predictions of Ref.~\cite{Calcietal16}.}
\begin{ruledtabular}
\begin{tabular}{ccccccccc}
$\sigma$ &$\beta$ & $V^{(0)}_{1/2^+}$ & $V^{(2)}_{1/2^+}$ & $E_{1/2^+}$ & ${\cal C}_{1/2^+}$ & ${\cal S}_{{1/2^+}}$ \\ 
(fm) & & (MeV) & (MeV fm$^{-2}$) & (MeV) & (fm$^{-1/2}$) & \\ \hline
1.3&0.35  & $-67.810$ & $-21.5$ & $-0.5031$& 0.7862   & 0.98  \\
1.5&0.5 & $-66.144$ & $-6.3$ & $-0.5031$ & 0.7862  & 0.92 \\
1.8&0.5 & $-67.256$  &1.30 & $-0.5031$ & 0.7860  & 0.87 \\
2.0&0.5 & $-62.365$  & 2.01 & $-0.5031$ & 0.7862 & 0.85 \\ \hline
& \textit{ab initio} & \cite{Calcietal16}  &   & $-0.5$ & 0.786 & 0.90  \\ 
\end{tabular}
\end{ruledtabular}
\end{table}

\begin{figure*}[ht] 
    \centering
    \begin{minipage}{0.50\linewidth}
        \centering
        \includegraphics[width=\linewidth]{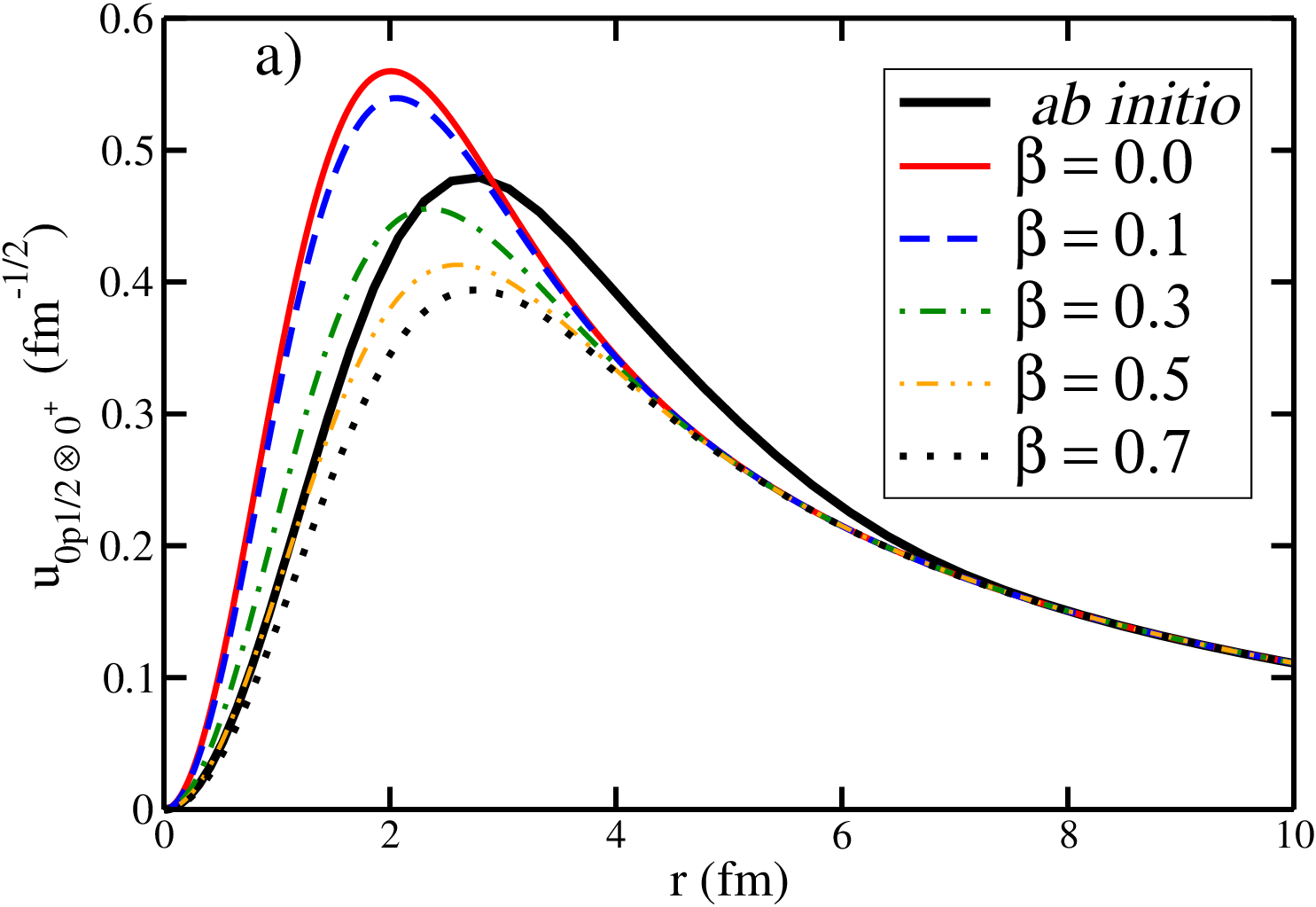}
    \end{minipage}\hfill
    \begin{minipage}{0.50\linewidth}
        \centering
        \includegraphics[width=\linewidth]{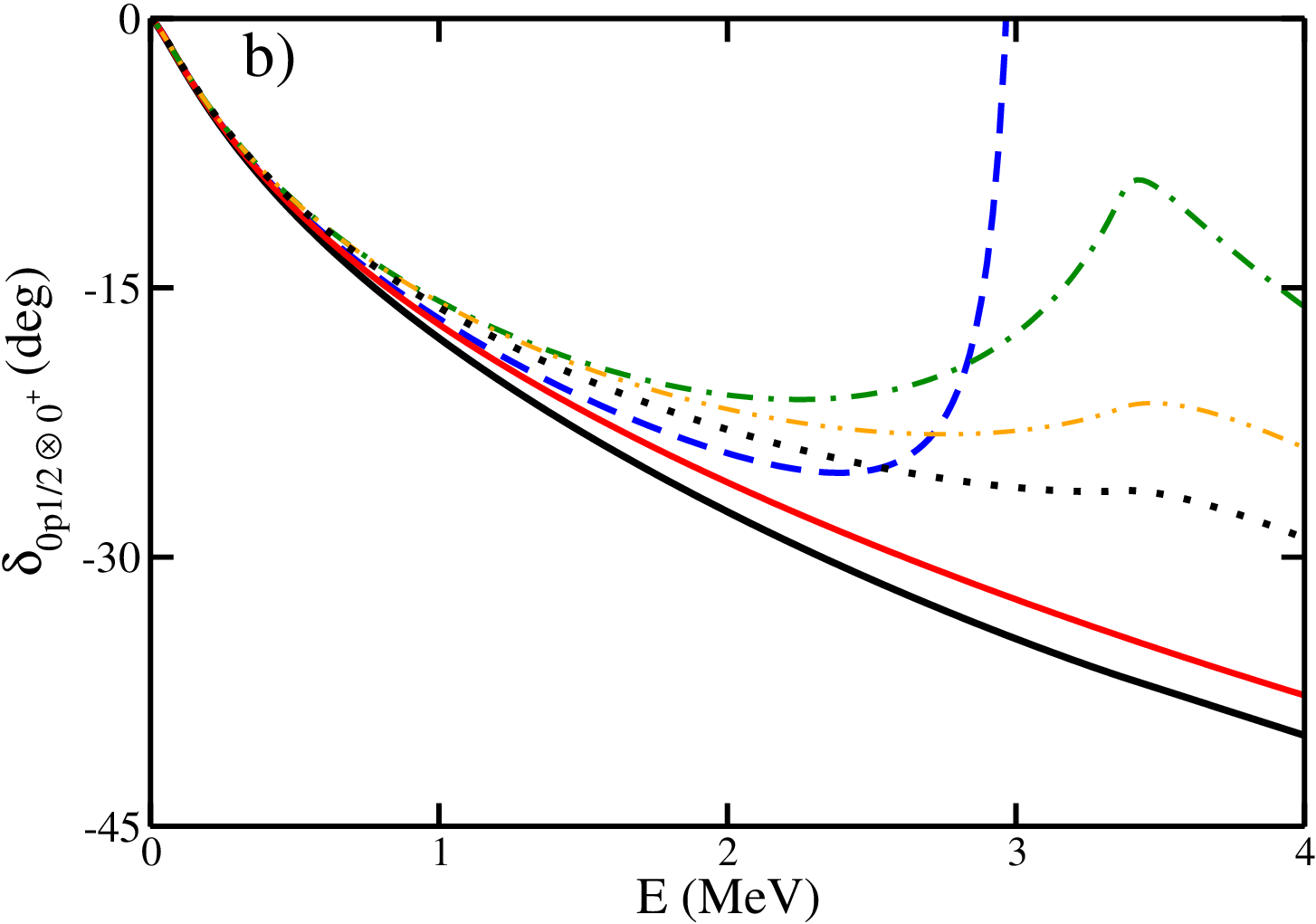}
    \end{minipage}
    \caption{Halo-EFT particle-rotor calculation of the $\frac{1}{2}^-$ ground state of $^{11}$Be for different values of the deformation parameter $\beta$ with $\sigma=1.5$~fm. (a) Radial wave functions in the $0p_{1/2} \otimes 0^+$ channel. (b) Corresponding eigenphaseshifts as a function of the $^{10}$Be-$n$ relative energy $E$.
    The \textit{ab initio} predictions are plotted in thick black line \cite{Calcietal16}.}
     \label{fig:es_wf_up_15_15}
\end{figure*}

As mentioned above, all values of $\sigma$ provide very similar results.
Including the coupling with $\beta\sim0.5$ ($\beta=0.35$ for $\sigma=1.3$~fm) leads to solutions very close to the \emph{ab initio} predictions (solid black line), independently of the cutoff $\sigma$.
This improvement over the usual Halo-EFT approach ($\beta=0$, dotted lines) is observed for both the overlap wave function and  the eigenphaseshifts in the $s_{1/2} \otimes 0^+$ channel. 

Including core excitation in the effective few-body model of $^{11}$Be has the desired effect of shifting the breakdown scale of the Halo-EFT up to larger energies.
This rotor-EFT leads to a good description of the \emph{ab initio} phaseshift up to, at least, 4~MeV, far beyond the 1.5~MeV of the usual Halo-EFT used in Ref.~\cite{Capeletal18}.
Therefore, it has a clear advantage compared to the latter to emulate \emph{ab initio} calculations at a relatively low computational cost.
This will enable us to improve the description of halo nuclei within nuclear-reaction models.

\subsection{Bound excited state: 1/2$^-$}\label{es}

We now turn to the $\frac{1}{2}^-$ bound excited state of $^{11}$Be.
In an extreme shell-model description, this state should be the ground state with six neutrons filling the $0s_{1/2}$ and $0p_{3/2}$ subshells leaving the unpaired valence neutron in the $0p_{1/2}$ subshell. However, because of the well-known shell inversion in $^{11}$Be, it corresponds to the first---and only---bound excited state.

Within the particle-rotor model described in Sec.~\ref{model}, that state is seen as the superposition of three possible configurations:
\begin{equation}
\ket{\frac{1}{2}^-}=\ket{0p_{1/2} \otimes 0^+} + \ket{0p_{3/2} \otimes 2^+} + \ket{0f_{5/2} \otimes 2^+}.
\label{overlap_wf_es}
\end{equation}

As for the ground state, we adjust the LECs $V^{(0)}_{1/2^-}$ and $V^{(2)}_{1/2^-}$ of the $c$-$n$ potential \eq{Vcn_NLO_central} to reproduce the experimental energy $E_{1/2^-}=-0.184$~MeV \cite{Ajzen1990} and the \textit{ab initio} predicted ANC ${\cal C}_{1/2^-}=0.129$~fm$^{-1/2}$ \cite{Calcietal16}.
The calculations have been performed for different values of $\beta$ to study the influence of this coupling upon the radial wave function of the dominant $0p_{1/2} \otimes 0^+$ channel and the corresponding eigenphaseshift $\delta_{p_{1/2} \otimes 0^+}$.
To estimate the role of short-range physics in this model, the cutoff of the NLO potential \eqref{Vcn_NLO_central} has been varied from $\sigma=1.2$, 1.3 1.5, 1.8, and 2.0~fm.
For brevity, we show results only for $\sigma=1.5$~fm.

The wave functions and eigenphaseshifts obtained in this scenario are displayed in Fig.~\ref{fig:es_wf_up_15_15}.
Table~\ref{tab:gs_wf_dep_sol1_v2up_1.5_2.2} lists, for each value of $\beta$, the LECs of the $c$-$n$ potential alongside the structure observables they produce.

\begin{table}[b]
\caption{\label{tab:gs_wf_dep_sol1_v2up_1.5_2.2}
Parameters of the effective $^{10}$Be-$n$ potential \eqref{Vcn_NLO_central} in the $\frac{1}{2}^-$ excited state with $\sigma=1.5$~fm.
For different $\beta$, we provide the LECs, which reproduce the listed $E_{1/2^-}$ energy, ANC, and SF.
The \textit{ab initio} predictions of Ref.~\cite{Calcietal16} are given in the last line.}
\begin{ruledtabular}
\begin{tabular}{cccccccc}
 $\beta$ & $V^{(0)}_{1/2^-}$ & $V^{(2)}_{1/2^-}$ & $E_{1/2^-}$ & ${\cal C}_{1/2^-}$ & ${\cal S}_{0p_{1/2} \otimes 0^+}$ \\ 
  & (MeV) & (MeV fm$^{-2}$) & (MeV) & (fm$^{-1/2}$) & \\ \hline
0.0  & $-83.625$ & 5.2  & $-0.1841$ & 0.12908 & 1.00  \\
0.1 & $-80.6825$ & 4.6 & $-0.1841$ & 0.12911  & 0.96 \\
0.3 & $-64.2042$ & 1.8 & $-0.1841$ & 0.12909  & 0.78 \\
0.5 & $-49.8437$  & 0.42 & $-0.1841$ & 0.12912 & 0.69 \\
0.7 & $-39.6605$  & 0.35 & $-0.1841$ & 0.12908 & 0.64  \\  \hline
\textit{ab initio} & \cite{Calcietal16}  &      & $-0.1848$ & 0.1291 &  0.85
\end{tabular}
\end{ruledtabular}
\end{table}

Since the LECs are fitted to the one-neutron separation energy and the \emph{ab initio} ANC of the excited state, all our solutions reproduce the asymptotics of the prediction of Calci \etal \cite{Calcietal16}.
Nevertheless, they differ significantly from that overlap wave function.
First, our radial wave functions reach their asymptotic behavior at $r\sim4$~fm, whereas the \emph{ab initio} one becomes fully asymptotic only at $r\gtrsim7$~fm.
Between 4 and 7 fm, in what we coin the \textit{pre-asymptotic} region, the microscopic calculation is significantly larger than our particle-rotor effective model.
This is independent of $\beta$: even with a large coupling, we cannot reproduce this feature of the NCSMC.

Second, our particle-rotor radial wave function peaks at a smaller radius than the \emph{ab initio} one.
The attractive nature of the coupling partially corrects that, for the radial wave function is shifted to larger radii with increasing $\beta$.
With a large coupling constant $\beta=0.7$, the position of the \emph{ab initio} wave function is properly reproduced.
However, at that value, the coupling is so strong that we underestimate the NCSMC spectroscopic factor by more than 20\%.
Therefore, unlike for the $\frac{1}{2}^+$ ground state, the coupling to the $2^+$ excited state of $^{10}$Be does not enable us to improve the Halo-EFT of Ref.~\cite{Capeletal18} to reproduce the \emph{ab initio} calculation.

This conclusion is confirmed in Fig.~\ref{fig:es_wf_up_15_15}(b), which shows the eigenphaseshifts for the dominant ${p_{1/2} \otimes 0^+}$ channel.
Not only does the coupling not improve the agreement between Halo-EFT and the \emph{ab initio} prediction, but it worsens it.
With just a small coupling $\beta=0.1$ (blue dashed line), a resonant-like behavior appears in the phaseshift.
That resonance is located at $E\approx 3.2$~MeV. 
This effect is due to an eigenstate of the $V_{cn}$ interaction in the ${0p_{3/2} \otimes 2^+}$ channel. Because the diagonal potential \eq{Vcn_NLO_central} is the same in all channels, we expect, for small $\beta$, to observe a state at about $\varepsilon_{0p_{1/2}}+\epsilon_{2^+}$ in that channel, where $\varepsilon_{n_r\ell j}$ is the energy of the single-particle bound state obtained in partial wave $\ell j$ by the potential $V$ \eq{Vcn_NLO_central} without coupling.
With increasing coupling, the resonance becomes less marked and shifts to higher energies.
However, no $\beta>0$ leads to an eigenphaseshift in agreement with the \emph{ab initio} prediction.

Contrary to the case of the $\frac{1}{2}^+$ ground state, the Halo-EFT description of the excited bound state of $^{11}$Be cannot be improved by including core excitation. Even exploring the model space up to large values of the cutoff $\sigma$, e.g., up to 2.4~fm, we find no set of LECs that leads to a fair description of the \emph{ab initio} prediction for both the overlap wave function and the eigenphaseshift \cite{KubushishiThesis24}.
As shown in \anx{essc}, shifting the coupling to larger radii does not help either.
The only way to improve the description of the $\frac{1}{2}^-$ excited state of $^{11}$Be seems to use a usual Halo-EFT description at N$^2$LO \cite{KC25}.

As explained in Ref.~\cite{SBE93}, the difference in the role of core excitation in the structure of the two bound states of $^{11}$Be is due to Pauli blocking.
In the negative-parity state, the configuration in which the core is in its $2^+$ excited state is blocked by the presence of the additional valence neutron in the $0p$ shell.
The structure of that $\frac{1}{2}^-$ is therefore not well described in the deformed cluster model considered here and in previous work \cite{VinhMau95,Nunes1996}, as already mentioned in Ref.~\cite{Esbensen1995}.

\subsection{Electric dipole transition probability} \label{E1}

The results shown in Secs.~\ref{gs} and \ref{es} offer a more detailed description of the structure of both bound states of $^{11}$Be than the Halo-EFT of Ref.~\cite{Capeletal18}.
To test this model on an observable, we compute the electric dipole transition probability between the bound states of $^{11}$Be: ${\cal B}(E1 ; \frac{1}{2}^+\rightarrow \frac{1}{2}^-)$.
This observable has been computed \textit{ab initio} \cite{Calcietal16} and measured experimentally \cite{measurementBE1lifetime,Nak97,be1_SUMMERS2007,kwan_exp1_be1}.

The reduced electric transition probability for a transition of multipolarity $\lambda$ from an initial bound state $\ket{J}$ to a final bound state $\ket{J'}$ reads
\begin{equation}
 \label{eq:be1LayMoro}
{\cal B}(E\lambda; J \to J')=\frac{2 J'+1}{2 J+1}\left | 
\langle J' \| \mathcal{M}(E\lambda) \| J \rangle      \right |^2    ,
\end{equation}
where $\ket{J}$ and $\ket{J'}$ are assumed to be normalized to unity, and $\mathcal{M}(E\lambda\mu)$ designates the electromagnetic multipole operator.
Following Ref.~\cite{Arai2010}, for a one-neutron halo nucleus, the electric dipole transition operator can be decomposed as follows
\beq
\mathcal{M}(E1 \mu)&= & \mathcal{M}_{c}(E 1 \mu)-\frac{Z}{A} e\ r\ Y_{1}^{\mu}(\hat{r}),
\eeqn{eq:be1}
where the first term of the right-hand side corresponds to the dipole operator of the core $c$, whereas the second term accounts for the recoil of the core.
Because the present model of $^{11}$Be includes only core states of the same parity, the former plays no role in the calculations.

The ${\cal B}(E1 ; \frac{1}{2}^+ \rightarrow \frac{1}{2}^-$) computed within this rotor-EFT are displayed in Table.~\ref{tab:BE1table}.
We use the same cutoff $\sigma=1.5$~fm and deformation parameter $\beta$ for both bound states. We consider the value of $\beta$ that best reproduces the \textit{ab initio} ground state wave function, see Fig.~\ref{fig:best1/2+_type1solution}.

\begin{table}[b]
\caption{\label{tab:BE1table}
Electric dipole transition probabilities ${\cal B}(E1;\frac{1}{2}^+ \rightarrow \frac{1}{2}^-)$ in our rotor-EFT obtained with the same cutoff and deformation for both bound states \cite{KubushishiThesis24}.
These values are compared to the \textit{ab initio} prediction \cite{Calcietal16} and the experimental data \cite{measurementBE1lifetime,Nak97,be1_SUMMERS2007,kwan_exp1_be1}.
The non-zero $\beta$ correspond to those that best describe the \textit{ab initio} ground state wave function (see Fig.~\ref{fig:best1/2+_type1solution}).}
\begin{tabular}{cccc}\hline\hline
 & & \multicolumn{2}{c}{${\cal B}(E1;\frac{1}{2}^+ \rightarrow \frac{1}{2}^-)$ (e$^2$fm$^2$)} \\ 
  $\sigma$ (fm)& & $\beta \neq 0$ & $\beta=0$ \\  \hline
1.3 & & 0.104 & 0.103\\ 
1.5 & & 0.106 & 0.106\\
1.8 & & 0.109 & 0.108 \\
2.0 & & 0.110 & 0.109\\\hline
\textit{ab initio} & \cite{Calcietal16}  & 0.117 \\ \hline 
Experiments & &\\
Millener \etal & \cite{measurementBE1lifetime} & 0.116(12) & \\
Nakamura \etal & \cite{Nak97} & 0.099(10) & \\ 
Summers \etal & \cite{be1_SUMMERS2007} & 0.105(12) & \\ 
Kwan \etal  & \cite{kwan_exp1_be1} & 0.102(2) & \\\hline\hline
\end{tabular}
\end{table}

All configurations lead to similar ${\cal B}(E1)$, which are slightly lower than the \emph{ab initio} prediction \cite{Calcietal16}.
The reason for that small underestimation is due to the \emph{pre-asymptotic} region of the $\frac{1}{2}^-$ excited bound state computed \emph{ab initio}, which none of our rotor-EFT calculations reproduce, see Fig.~\ref{fig:es_wf_up_15_15}(a).
The larger overlap between both radial wave functions obtained with that feature explains the small difference.
Using larger values of $\sigma$ slightly reduces that difference \cite{KubushishiThesis24}.

Our calculations are also in good agreement with data \cite{measurementBE1lifetime,Nak97,be1_SUMMERS2007,kwan_exp1_be1}.
However, they are on the higher end of the experimental uncertainty range of the most recent---and most precise---experiment \cite{kwan_exp1_be1}.
It is therefore possible that the pre-asymptotic behavior observed in the \emph{ab initio} calculation is not physically meaningful.

Table~\ref{tab:BE1table} lists also the ${\cal B}(E1)$ obtained without core excitation ($\beta=0$).
Because these values are identical to those with $\beta\ne0$, we conclude that this observable is independent of the core excitation and that it is a peripheral observable, which probes only the $S_n$ and ANC.

\section{Conclusions and outlook}\label{conclusion}
Halo-EFT is very efficient to describe halo nuclei within reaction models \cite{Capeletal18,MC19,MYC19,CP23,YC18,HC19,HC21}.
Its order-by-order expansion enables us to study systematically how nuclear structure affects reaction observables.
This expansion is of course limited by its breakdown scale.
Structure properties that lie beyond it cannot be explicitly accounted for.
In particular, the core excitation is overlooked in usual Halo-EFT \cite{bertulani2002review,Hammer17review}, although it has been shown to play a role in the breakup of $^{11}$Be on carbon \cite{ML12}.
In this work, we develop and study an extension of Halo-EFT that includes the excitation of the core.

Following previous work \cite{Esbensen1995,Nunes1996,ThompsonFaCE}, we describe the core as a rigid rotor to which a neutron is bound by an effective potential.
Truthful to Halo-EFT, we consider that interaction to have the shape of a Gaussian of range $\sigma$ and its derivative.
The coupling between the core states is treated perturbatively.

We apply this model to $^{11}$Be, the archetypical one-neutron halo nucleus, which has been computed \emph{ab initio} within NCSMC \cite{Calcietal16}.
This provides us with an accurate description of the nucleus, upon which the LECs of our rotor-EFT can be adjusted.
In this first study, we consider only the first two states of the $^{10}$Be core: its $0^+$ ground state and its first $2^+$ excited state.
We vary the deformation parameter $\beta$ to analyze in detail the sensitivity of the calculations to the core excitation.

Including core excitation within the Halo-EFT description of the $\frac{1}{2}^+$ ground state of $^{11}$Be improves the agreement with the \emph{ab initio} prediction of Calci \etal \cite{Calcietal16}.
With a value of the deformation $\beta\sim0.5$ nearly independent of the cutoff, we obtain a radial wave function in the dominant $1s_{1/2}\otimes 0^+$ channel very close to the NCSMC overlap wave function.
Moreover, the corresponding eigenphaseshift agrees with the \emph{ab initio} prediction up to 4~MeV, which is a significant gain compared to the usual Halo-EFT \cite{Capeletal18}.
This extension of the breakdown scale of our effective theory enables us to better emulate the NCSMC calculations within a computationally affordable few-body description of $^{11}$Be.
Such a description is very valuable in nuclear-reaction theory to test \emph{ab initio} predictions on reaction observables.

Unfortunately, the results in the $\frac{1}{2}^-$ excited state are less good.
Independently of the cutoff $\sigma$ and the deformation parameter $\beta$, we do not find a solution of our rotor-EFT that improves the agreement with the \emph{ab initio} prediction for both the overlap wave function and the phaseshift in the $p_{1/2}\otimes 0^+$ channel. This issue is due to Pauli blocking, which hinders the population of the $0p_{3/2}\otimes 2^+$ in that state \cite{SBE93}.
Since that effect is not accounted for within our rotor-EFT description of $^{11}$Be, it is no surprise, that we fail to describe that mean-field state  \cite{Calcietal16,PEI200629,Esbensen1995}.

Despite the imperfect description of the excited state, we have computed the electric dipole transition probability between both bound states ${\cal B}(E1;\frac{1}{2}^+ \rightarrow \frac{1}{2}^-)$.
Using the parameters providing the best agreement with the \emph{ab initio} prediction for the ground state, we obtain values in good agreement with experiment \cite{measurementBE1lifetime,Nak97,be1_SUMMERS2007,kwan_exp1_be1}. We note that the coupling has no significant effect on the calculations confirming that this observable is peripheral and probes only the energy and ANC of the $^{11}$Be bound states.

These results show that it is not only possible to include the core degree of freedom within Halo-EFT, but that it enables us to improve that description, when we compare it to a fully \emph{ab initio} approach.
Being obtained at a relatively low computational cost, it can be included in reaction models. In a parallel work \cite{KC25c}, we extend this study to the resonant continuum of $^{11}$Be, which is key to compute breakup cross sections \cite{Fuk04,CGB04,ML12,CPH22}.
This will enable us to study in detail the influence of core excitation on breakup reactions.
Moreover, the higher breakdown scale of this new rotor-EFT will enable us to test the predictions of accurate structure models on experimental data at a finer scale.
Finally, this formalism could be used to study other one-neutron halo nuclei, for which core excitation seems to play an important role, such as $^{19}$C \cite{LayCarbon19} or $^{31}$Ne \cite{deformationNe31}.

\begin{figure*}[ht] 
    \centering
    \begin{minipage}{0.50\linewidth}
        \centering
        \includegraphics[width=\linewidth]{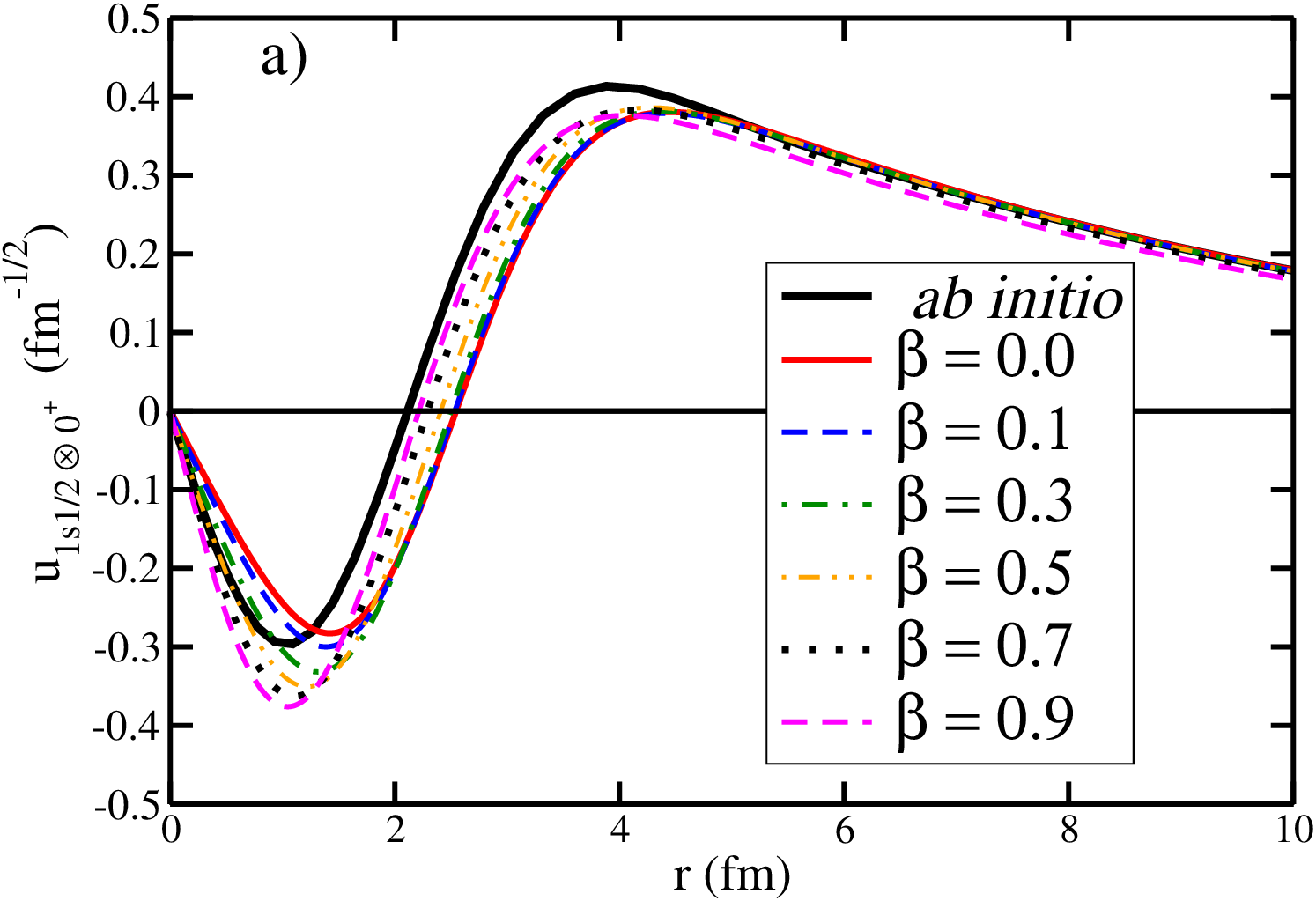}
    \end{minipage}\hfill
    \begin{minipage}{0.50\linewidth}
        \centering
        \includegraphics[width=\linewidth]   {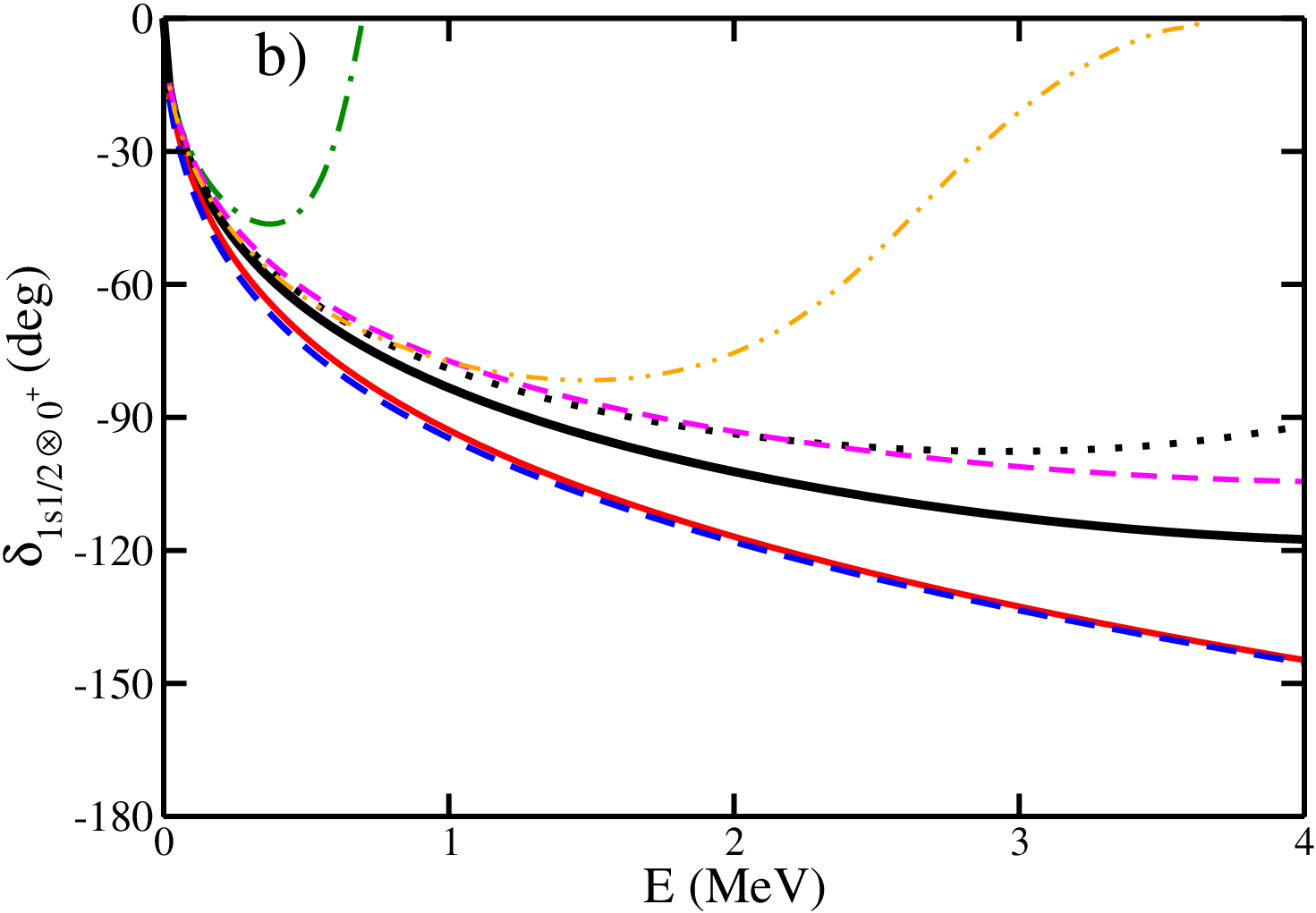}
    \end{minipage}
    \caption{
    Second solution for the $\frac{1}{2}^+$ ground state of $^{11}$Be within the Halo-EFT particle-rotor model. The results are shown for different values of the deformation parameter $\beta$ using $\sigma=1.5$~fm. (a) Radial wave functions in the $1s_{1/2} \otimes 0^+$ channel. (b) Corresponding eigenphaseshifts as a function of the $^{10}$Be-$n$ relative energy $E$.
    The \textit{ab initio} predictions are plotted in thick black lines \cite{Calcietal16}.
    }
    \label{fig:gs_wf_down_15_15}
\end{figure*}

\begin{acknowledgments}
We thank P. Descouvemont for insightful discussions on the R-Matrix method, D. R. Phillips, H.-W. Hammer and G. Singh for interesting discussions and comments. This work was supported by the Deutsche Forschungsgemeinschaft (DFG, German Research Foundation) – Project-ID 279384907 – SFB 1245 and performed in part under the auspices of the U.S.~Department of Energy under contract No.~DE-FG02-93ER40756
\end{acknowledgments}

\appendix

\section{\textbf{Another description of the $1/2^+$ ground state: A solution with a near degeneracy}}\label{type2}

Scanning the parameter space more broadly, we have discovered another solution for the $\frac{1}{2}^+$ ground state.
It arises from a close degeneracy between the dominant $1s_{1/2}$ bound state in the diagonal $c$-$n$ potential and $0d_{5/2}$ and $0d_{3/2}$ bound states, which appears for a particular set of the LECs $V^{(0)}_{1/2^+}$ and $V^{(2)}_{1/2^+}$.
This second solution is of little physical interest in the study of $^{11}$Be.
First, such a degeneracy has no physical ground.
Second, as we show below, it worsens the agreement between Halo-EFT and the \emph{ab initio} predictions.
Nevertheless, for completeness, we detail this unexpected case here.

Although it is not fully surprising that in such coupled-channel problem, two couples of LECs lead to the desired binding energy and ANC, this second solution is strikingly different from the efficient results of \Sec{gs}.
This can be seen by comparing Fig.~\ref{fig:gs_wf_down_15_15} and Table~\ref{tab:gs_wf_down_15_15} of this Appendix with Fig.~\ref{fig:gs_wf_up_15_15} and Table~\ref{tab:gs_wf_up_15_15} of \Sec{gs}.

\begin{table}[b]
\caption{\label{tab:gs_wf_down_15_15}
Second solution for the $\frac{1}{2}^+$ ground state. Parameters of the effective $^{10}$Be-$n$ potential \eqref{Vcn_NLO_central} for $\sigma=1.5$~fm.
For each deformation parameter $\beta$, we provide the depths V$^{(0)}_{1/2^+}$ and V$^{(2)}_{1/2^+}$, which reproduce the listed $E_{1/2^+}$ energy, ANC, and SF.
The \textit{ab initio} predictions of Ref.~\cite{Calcietal16} are given in the last line.}
\begin{ruledtabular}
\begin{tabular}{cccccccc}
 $\beta$ & $V^{(0)}_{1/2^+}$ & $V^{(2)}_{1/2^+}$ & $E_{1/2^+}$ & ${\cal C}_{1/2^+}$ & ${\cal S}_{1s_{1/2} \otimes 0^+}$ \\ 
  & (MeV) & (MeV fm$^{-2}$) & (MeV) & (fm$^{-1/2}$) & \\ \hline
0 & $-7.771$ & $-27.192776$  & $-0.5031$ & 0.7861 & 0.80 \\
0.1 & $-12.409$ & $-25.45293$ & $-0.5031$ & 0.7863  & 0.82 \\
0.3 & $-20.881$ & $-21.960$ & $-0.5031$ & 0.7863  & 0.85 \\
0.5 & $-32.7285$ & $-17.000$ & $-0.5031$  & 0.7857 & 0.88 \\
0.7 & $-46.195$ & $-11.000$ & $-0.5031$  & 0.7646 & 0.87\\
0.9 & $-55.6975$ & $-6.000$ & $-0.5031$  & 0.7382 & 0.85\\ \hline
\textit{ab initio} &   &   & $-0.5$ & 0.786 & 0.90  \\ 
\end{tabular}
\end{ruledtabular}
\end{table}

As in the first solution, all the reduced radial wave functions $u_{1s_{1/2} \otimes 0^+}$ of this second solution exhibit the same asymptotics.
The coupling induced by the deformation affects only the internal part of the wave function.
However, in this case, the node is located at a larger radius than the \emph{ab initio} one.
Introducing the coupling leads to a progressive evolution of our solution towards the overlap wave function of Calci \etal \cite{Calcietal16}.
Nevertheless, even at large $\beta$, we never reach as good an agreement with the \emph{ab initio} prediction as in Fig.~\ref{fig:gs_wf_up_15_15}(a) for $\beta=0.5$.

The main reason for this additional solution is the existence of $0d_{5/2}$ and $0d_{3/2}$ bound states in the vicinity of the dominant $1s_{1/2}$ in the mean field generated by the diagonal term of $V$ \eq{Vcn_NLO_central}.
This is unsurprising in the usual shell-model description of nuclei since that intruder state is within the $sd$ shell.
Because the effective $c$-$n$ interaction \eq{Vcpl_order1} contains no spin-orbit coupling term, a particular set of LECs can lead to a degeneracy at $\beta=0$ in the $1s_{1/2}\otimes 0^+$, $0d_{5/2}\otimes 2^+$, and $0d_{3/2}\otimes 2^+$ channels, viz.\ $\varepsilon_{1s1/2}=\varepsilon_{0d5/2}+\epsilon_{2^+}=\varepsilon_{0d3/2}+\epsilon_{2^+}$, where $\varepsilon_{n_r\ell j}$ is the energy of the single-particle bound state obtained in partial wave $\ell j$ by the potential $V$ \eq{Vcn_NLO_central} without coupling. This effect corresponds to what has been presented in Sec.~\ref{es}, when a single-particle eigenstate in the $p_{3/2}\otimes 2^+$ channel lead to a resonance in the $p_{1/2}\otimes 0^+$ eigenphaseshift.
It has also been observed within a more traditional particle-rotor model of $^{11}$Be in Ref.~\cite{CDN10}.
In that case, even without coupling, the solution can exhibit a spectroscopic factor lower than 1, see the first line of Table~\ref{tab:gs_wf_down_15_15}; see also Figs.~4(a) and 5(a) of Ref.~\cite{CDN10}.
All things being equal, a drop in ${\cal S}_{1s_{1/2} \otimes 0^+}$ leads to a reduction of the ANC ${\cal C}_{1/2^+}$.
To counter this effect, it is necessary to attract the wave function further outwards, viz.\ deepen the surface term of $V_{cn}$.
Then, to keep the one-neutron separation energy fixed, the central term needs to be shallower.
This explain qualitatively the values of the LECs provided in the first line of Table~\ref{tab:gs_wf_down_15_15}.

When the coupling is introduced on that second solution with $\beta=0$, the LECs of the $c$-$n$ interaction evolves as shown in Table~\ref{tab:gs_wf_down_15_15}.
This evolution explains the changes observed in Fig.~\ref{fig:gs_wf_down_15_15}(a).
The deepening of the central term, and the reduction of the attraction in the surface term lead to the shift of the radial wave function to smaller radii.

The inadequacy of this second solution to reproduce the \emph{ab initio} calculations is even clearer in the panel (b) of Fig.~\ref{fig:gs_wf_down_15_15}, which displays the corresponding eigenphaseshift. 
Contrary to the results of \Sec{gs}, none of these solutions provide a fair agreement with the \emph{ab initio} prediction (thick black line).
This is particularly surprising at low energy because all solutions exhibit the correct binding energy and ANC of the bound state.
The $S$ matrix therefore exhibits a pole at the right energy with the right residue \cite{VAANDRAGER2019}.
Following Ref.~\cite{JMScapel2010} and the results presented in \Sec{gs}, we would naively expect that the LECs of Table~\ref{tab:gs_wf_down_15_15} lead to the correct scattering length and effective range.
However it is by no means the case, showing that the single-channel results or Ref.~\cite{JMScapel2010} cannot be blindly transposed in a multi-channel model.
The understanding of this unexpected behavior requires further analyses, which are beyond the scope of this work.
Nevertheless, the aforementioned degeneracy is probably the reason for this disagreement, for it produces additional poles in the $S$ matrix in both $0d_{5/2}\otimes 2^+$ and $0d_{3/2}\otimes 2^+$ channels. 
This is confirmed when the coupling is introduced, which raises the degeneracy.
In particular, for $\beta=0.3$ (green dash-dotted line) and 0.5 (orange dash-dot-dotted line), we observe a resonant behavior, which can be related to the shift of the bound states in the $0d_j\otimes 2^+$ channels when $\beta$ is large enough.
The shallower surface term leads to higher single-particle energies $\varepsilon_{0d5/2}$ and $\varepsilon_{0d3/2}$.
At $\beta=0.3$ and 0.5, these energies appear in the continuum of the $1s_{1/2}\otimes 0^+$ channel, hence as resonances, as seen in Fig.~\ref{fig:es_wf_up_15_15}(b) for the $p_{1/2}\otimes 0^+$ channel of the excited state.

As the deformation parameter becomes larger, this resonance gets broader and is pushed further up in the continuum, beyond 4~MeV, so that for $\beta\ge0.7$, no resonant structures is observed.
At these larger deformations, the corresponding eigenphaseshifts provide a fairer agreement with the \textit{ab initio} prediction.
However, they never come close to the quality of the $\beta=0.5$ results shown in Fig.~\ref{fig:gs_wf_up_15_15}.
Moreover, they come at the cost of a large $\beta$ which is not consistent with the perturbative expansion used in \Eq{Vcpl_order1}. 
For this reason, and the fact that a bound state in neither $0d_{j}\otimes 2^+$ channels makes sense, this solution is not physically meaningful.

Note that, as in the solution presented in \Sec{gs}, we experience difficulties to accurately fit both the binding energy and the ANC of the wave function at large deformation.
At these $\beta$, the LECs of both solutions are close to each other.
The vicinity of these two solutions in our model space explains that even small changes in the potential depths can quickly switch our fitting code from one solution to the other.

There is no such solution for the $\frac{1}{2}^-$ excited state.
Because the diagonal potential $V$ \eq{Vcn_NLO_central} is the same in all partial waves, the single-particle state in the ${0p_{3/2} \otimes 2^+}$ is located at $\varepsilon_{0p_{1/2}}+\epsilon_{2^+}$, viz. in the continuum as seen in Fig.~\ref{fig:es_wf_up_15_15}(b).
Accordingly, no degeneracy can be obtained in this model.
It was however observed in Ref.~\cite{CDN10} using a $c$-$n$ interaction that includes a spin-orbit coupling term.
This term lowers the $0p_{3/2}$ orbitals compared to the $0p_{1/2}$, and can lead to $\varepsilon_{0p_{3/2}}+\epsilon_{2^+}\sim \varepsilon_{0p_{1/2}}$.

\section{Shifting the coupling to larger radii}\label{s/=sc}

The results presented in \Sec{es} show that the simplest version of the rotor EFT developed in \Sec{model} cannot reproduce the \emph{ab initio} description of the $\frac{1}{2}^-$ excited state of $^{11}$Be \cite{Calcietal16}.
In particular, introducing the coupling cannot explain the increase of the \emph{ab initio} overlap wave function in the \emph{pre-asymptotic} region at $4~{\rm fm}\lesssim r \lesssim 7$~fm, see Fig.~\ref{fig:es_wf_up_15_15}(a).
A tentative explanation is the relatively short range of the $c$-$n$ interaction considered in this EFT approach.
As mentioned in \Sec{results}, the range of the Gaussians potential \eq{Vcn_NLO_central} is chosen to be $\sigma\in[1.2,2.0]$~fm, which is smaller than the rms radius of the $^{10}$Be core $R_c\sim2.2$--2.4~fm \cite{Tan85l,Arai2010,descouvemont97}.
Thanks to its attractive nature, opting for a coupling located at a larger radius might help shift the radial wave function towards larger $c$-$n$ distances.
Moreover, this would be more in line with usual particle-rotor models, where the coupling term acts at the surface of the nucleus \cite{bohrmottelson,Esbensen1995,Nunes1996,ThompsonFaCE}.

To test this hypothesis, we perform series of calculations with a $c$-$n$ interaction, in which the coupling has a range $\sigma_c$ different from the diagonal term \eq{Vcn_NLO_central}.
In this extension of the model, \Eq{Vcpl_order1} reads
\begin{equation}
    V_{cn}(\ve{r},\xi)= V(r;\sigma) + \beta\, \sigma_c\, Y_{2}^{0}(\hat{r}') \frac{d}{d\sigma_c}V(r;\sigma_c).
    \label{eB1}
\end{equation}
While keeping $\sigma$ in the initial 1.3--2.0~fm range, we consider $\sigma_c=2.2$~fm.
That value corresponds to the calculated $^{10}$Be matter radius from Refs.~\cite{Arai2010,descouvemont97}, and is in fair agreement with the $^{10}$Be rms radius used in Refs.~\cite{Esbensen1995,Nunes1996}.

To see the influence such a choice has on the model, we repeat the calculations for both the ground and excited states presented in \Sec{results}.
We also check what happens in the non-physical solution presented in \anx{type2}.

\begin{figure*}[ht] 
    \centering
    \begin{minipage}{0.50\linewidth}
        \centering
        \includegraphics[width=\linewidth]
        {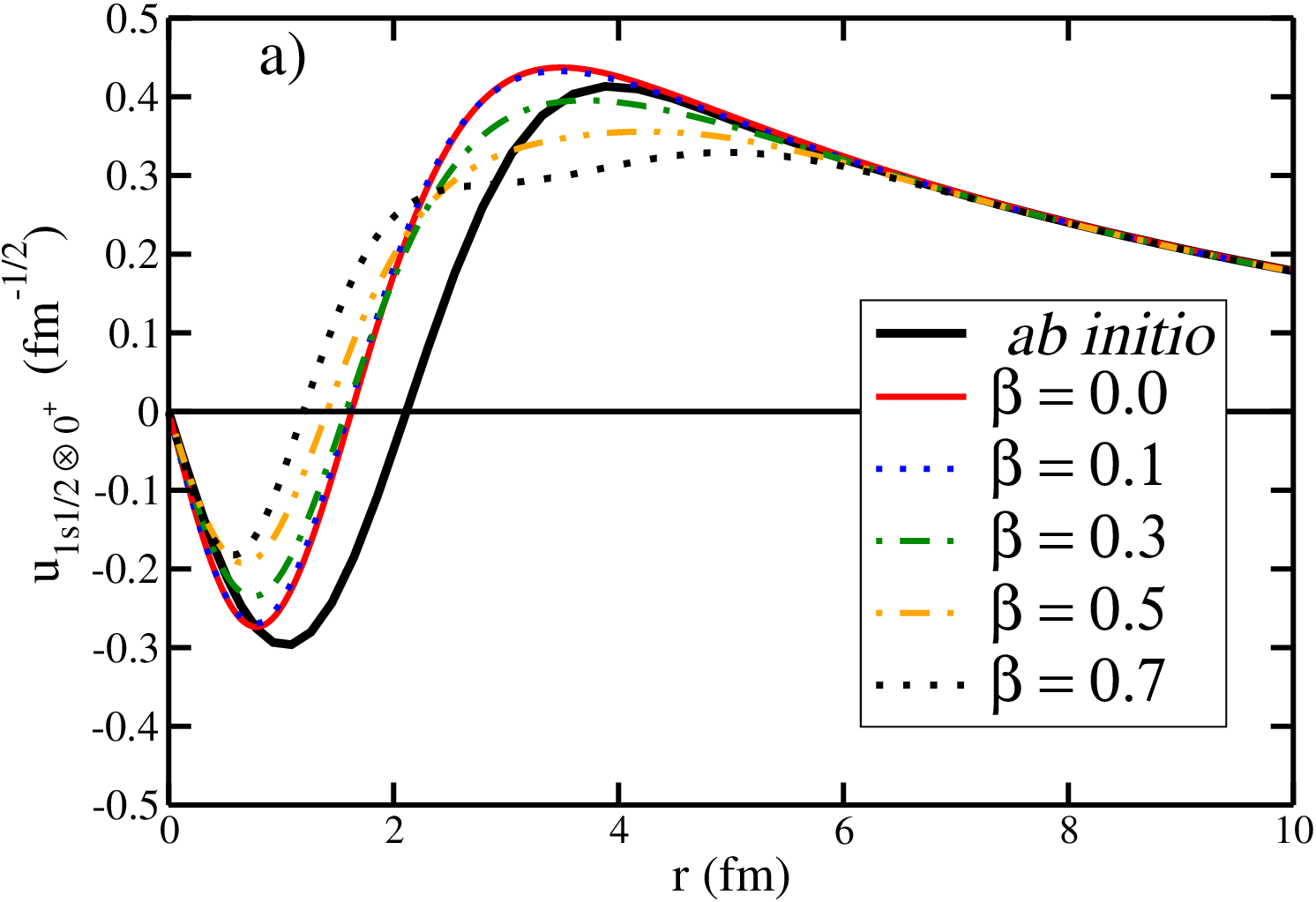}
    \end{minipage}\hfill
    \begin{minipage}{0.50\linewidth}
        \centering
        \includegraphics[width=\linewidth]
        {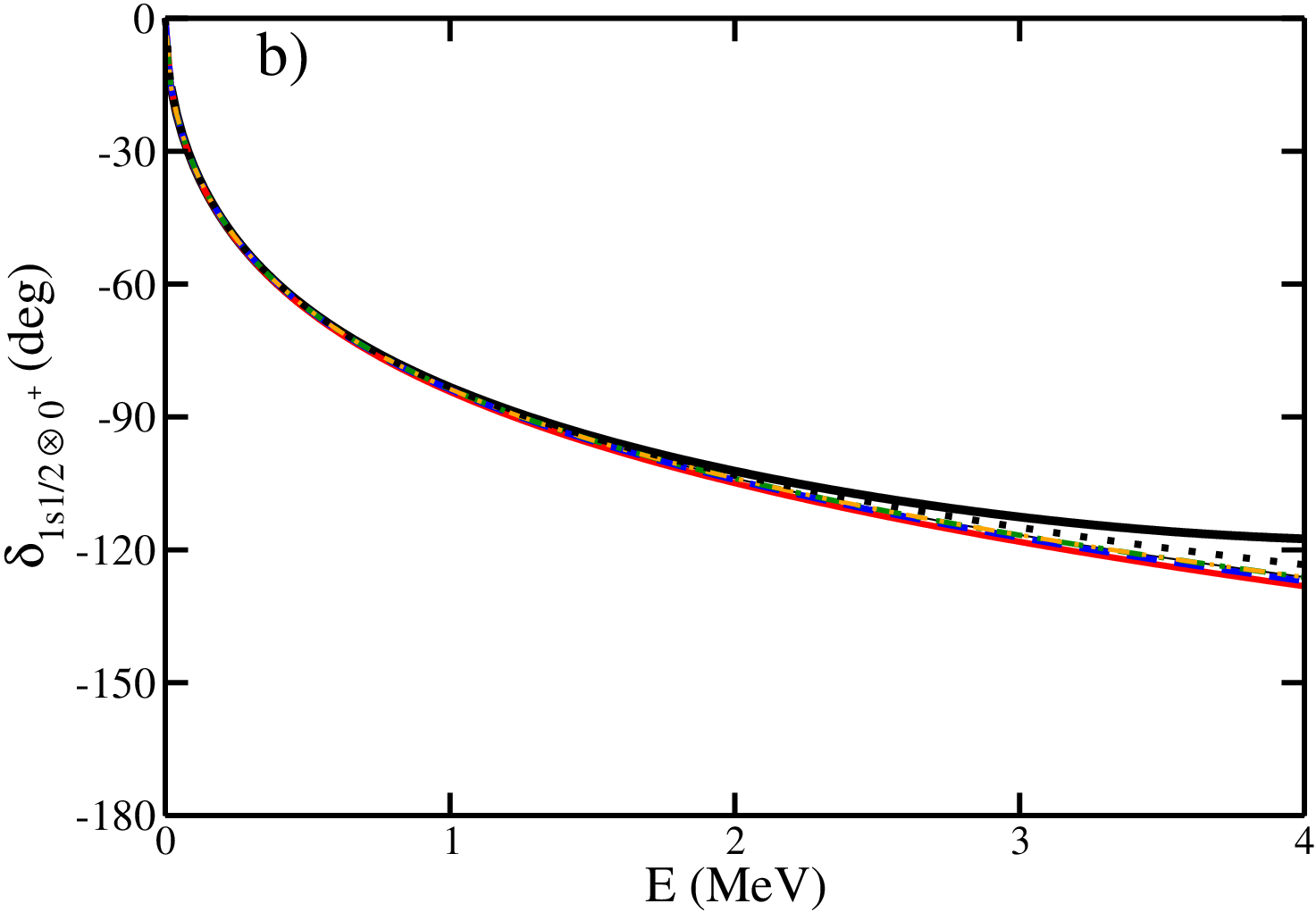}
    \end{minipage}
    \caption{Description of $^{11}$Be $\frac{1}{2}^+$ ground state in the extended rotor-EFT model of \Eq{eB1} with a coupling at $\sigma_c=2.2$~fm and a range of the diagonal potential of $\sigma=1.5$~fm.
    Solution without degeneracy, i.e., similar to Fig.~\ref{fig:gs_wf_up_15_15}.
    (a) Radial wave functions of $^{11}$Be $\frac{1}{2}^+$ ground state, in its $1s_{1/2} \otimes 0^+$ channel, computed for different values of the deformation parameter $\beta$. (b) Corresponding eigenphaseshifts in the $s_{1/2} \otimes 0^+$ channel as a function of the $^{10}$Be-$n$ scattering energy $E$. The \textit{ab initio} predictions of Ref.~\cite{Calcietal16} are also plotted in thick black lines.}
    \label{fig:gs_wf_up_15_22}
\end{figure*}

\subsection{Ground state: $1/2^+$}\label{gssc}

\subsubsection{\textbf{Solution without degeneracy}}\label{type1sc}

The results for the case without degeneracy, i.e., similar to what has been presented in \Sec{gs}, are illustrated in Fig.~\ref{fig:gs_wf_up_15_22} for $\sigma=1.5$~fm; other values of the range of the diagonal interaction \eq{Vcn_NLO_central} lead to similar results.
As before, for each value of $\beta$, both LECs of that interaction are fitted to reproduce the experimental one-neutron separation energy and the ANC predicted \emph{ab initio}, see Table~\ref{tab:gs_wf_dep_1.5_2.2_nodegeneracy}.
Accordingly, all the reduced radial wave functions in the dominant $1s_{1/2} \otimes 0^+$ channel exhibit the same asymptotics, see Fig~\ref{fig:gs_wf_up_15_22}(a).

\begin{table}[b]
\caption{\label{tab:gs_wf_dep_1.5_2.2_nodegeneracy} 
Shifting the coupling to larger radius. Parameters of the effective $^{10}$Be-$n$ potential \eqref{eB1} in the $\frac{1}{2}^+$ ground state with $\sigma_c=2.2$~fm while the range of the central interaction is kept at $\sigma=1.5$~fm.
For each deformation parameter $\beta$, we provide the depths V$^{(0)}_{1/2^+}$ and V$^{(2)}_{1/2^+}$, which reproduce the listed $E_{1/2^+}$ energy, ANC, and SF.
The \textit{ab initio} predictions of Ref.~\cite{Calcietal16} are given in the last line.
}
\begin{ruledtabular}
\begin{tabular}{cccccccc}
 $\beta$ & $V^{(0)}_{1/2^+}$ & $V^{(2)}_{1/2^+}$ & $E_{1/2^+}$ & ${\cal C}_{1/2^+}$ & ${\cal S}_{1s_{1/2} \otimes 0^+}$ \\ 
  & (MeV) & (MeV fm$^{-2}$) & (MeV) & (fm$^{-1/2}$) & \\ \hline
0 & $-100.190$ & 0  & $-0.5031$ & 0.7865 & 1.00 \\
0.1 & $-103.650$ & 1.15 & $-0.5031$ & 0.7860  & 0.99 \\
0.3 & $-109.058$ & 5.1 & $-0.5031$ & 0.7859  & 0.90 \\
0.5 & $-136.704$ & 15.2 & $-0.5031$  & 0.7863 & 0.82 \\
0.7 & $-183.289$ &29.5 & $-0.5031$  &0.7864 & 0.76\\ \hline
\textit{ab initio} & \cite{Calcietal16}  &   & $-0.5$ & 0.786 & 0.90  \\ 
\end{tabular}
\end{ruledtabular}
\end{table}

Unexpectedly, contrary to the original case where $\sigma_c=\sigma$, we could not find any set of LECs that reproduces the \emph{ab initio} overlap wave function.
No value of $\beta$ leads to a fair agreement with the microscopic calculation of Calci \etal \cite{Calcietal16}.
Worst, including the coupling reduces the agreement between our rotor-EFT calculation and the \emph{ab initio} overlap wave function: the node of our solution moves to smaller radius when $\beta$ increases.
Moreover, at $\beta=0.7$, we observe a local minimum between 3 and 4~fm.

This evolution of the rotor-EFT radial wave function with increasing coupling can be qualitatively related to the radial range, where it acts, and its global attractive effect.
The former characteristic implies that unlike in the $\sigma_c=\sigma$ case, the radial wave function reaches its asymptotic behavior at larger radius.
In  Fig.~\ref{fig:gs_wf_up_15_22}(a) this happens only at $r\sim 6$~fm, instead of around $r\sim4$~fm in Fig.~\ref{fig:gs_wf_up_15_15}(a).
The additional attraction provided by the coupling leads to a repulsive surface term of the diagonal potential to fit the ANC to its desired value (see Table~\ref{tab:gs_wf_dep_1.5_2.2_nodegeneracy}).
The binding energy is then obtained by deepening the central term.
This alternation of attractive-repulsive-attractive terms in the potential $V_{cn}$ explains the evolution of the radial wave function seen in Fig.~\ref{fig:gs_wf_up_15_22}(a).

The effect of the coupling on the eigenphaseshift remains modest, see Fig.~\ref{fig:gs_wf_up_15_22}(b).
Although the additional attraction induced by the coupling to the $0d_{5/2}\otimes 2^+$ and $0d_{3/2}\otimes 2^+$ channels slightly increases the phaseshift, the agreement with the \emph{ab initio} prediction never reaches the quality of the results obtained in Fig.~\ref{fig:gs_wf_up_15_15}(b) for $\beta=0.5$. 
Similar results are obtained with other values of $\sigma$, indicating the generality of this analysis.
This way to include core excitation within Halo-EFT is rather unsatisfactory.

\subsubsection{\textbf{Solution with a degeneracy}}\label{type2sc}

\begin{figure*}[ht] 
    \centering
    \begin{minipage}{0.50\linewidth}
        \centering
        \includegraphics[width=\linewidth]{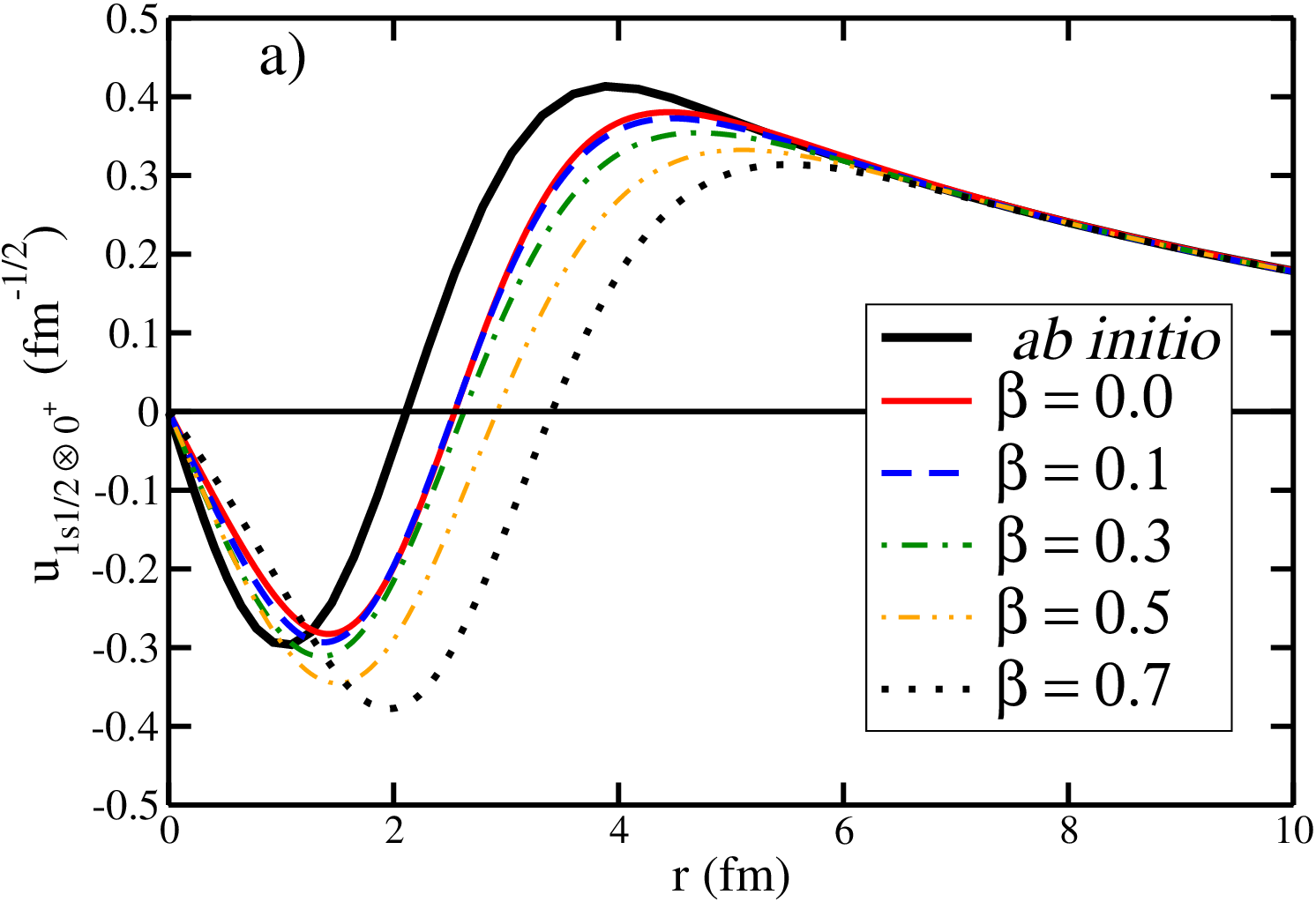}
    \end{minipage}\hfill
    \begin{minipage}{0.50\linewidth}
        \centering
        \includegraphics[width=\linewidth]{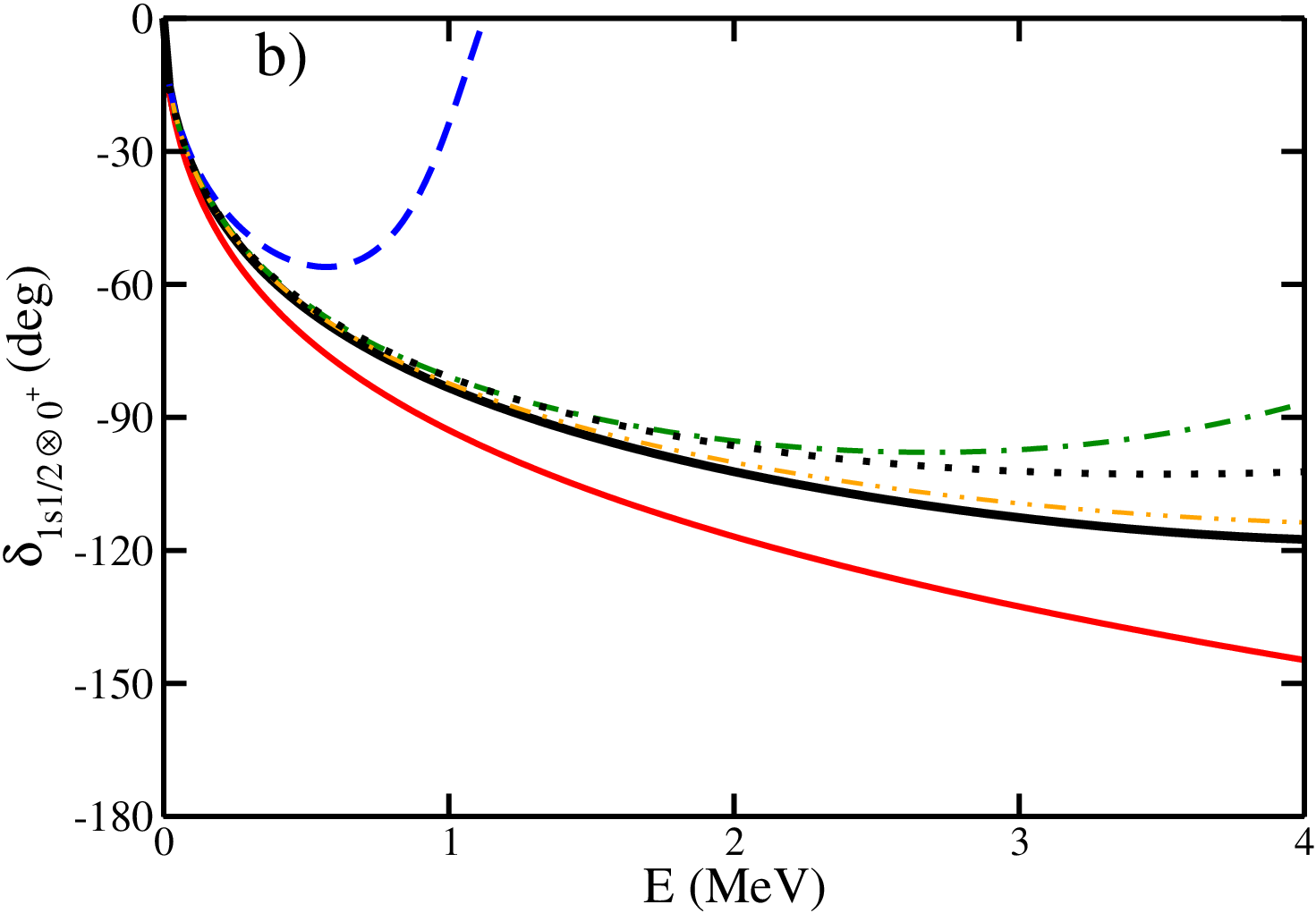}
    \end{minipage}
    \caption{Description of $^{11}$Be $\frac{1}{2}^+$ ground state in the extended rotor-EFT model of \Eq{eB1} with a coupling at $\sigma_c=2.2$~fm and a range of the diagonal potential of $\sigma=1.5$~fm.
    Solution with a degeneracy, i.e., similar to Fig.~\ref{fig:gs_wf_down_15_15}.
    (a) Radial wave functions of $^{11}$Be $\frac{1}{2}^+$ ground state, in its $1s_{1/2} \otimes 0^+$ channel, computed for different values of the deformation parameter $\beta$. (b) Corresponding eigenphaseshifts in the $s_{1/2} \otimes 0^+$ channel as a function of the $^{10}$Be-$n$ scattering energy $E$. The \textit{ab initio} predictions of Ref.~\cite{Calcietal16} are also plotted in thick black lines.}
    \label{fig:gs_wf_down_15_22}
\end{figure*}

In this case as well, exploring the parameter space has enabled us to find a set of LECs of the $c$-$n$ diagonal potential \eq{Vcn_NLO_central} that leads to degenerate bound states in the $1s_{1/2}\otimes0^+$ channel and the $0d_{5/2}\otimes 2^+$ and $0d_{3/2}\otimes 2^+$ channels, see Table~\ref{tab:gs_wf_dep_1.5_2.2_withdegeneracy}.
As in the previous section, the coupling does not lead to a better agreement of our radial wave function with the \emph{ab initio} one, see Fig.~\ref{fig:gs_wf_down_15_22}(a).
Here too, the fitted effective potential shifts the node of the rotor-EFT wave function away from the \emph{ab initio} prediction's. 
The additional attraction due to the coupling located at $\sigma_c=2.2$~fm $>\sigma$ drags the solution towards even larger radii.

\begin{table}[b]
\caption{\label{tab:gs_wf_dep_1.5_2.2_withdegeneracy} Second solution for the $\frac{1}{2}^+$ ground state obtained with the coupling shifted to larger radii. Parameters of the effective $^{10}$Be-$n$ potential \eqref{eB1} with $\sigma_c=2.2$~fm and $\sigma=1.5$~fm.
For each deformation parameter $\beta$, we provide the depths V$^{(0)}_{1/2^+}$ and V$^{(2)}_{1/2^+}$, which reproduce the listed $E_{1/2^+}$ energy, ANC, and SF.
The \textit{ab initio} predictions of Ref.~\cite{Calcietal16} are given in the last line.
}
\begin{ruledtabular}
\begin{tabular}{cccccccc}
 $\beta$ & $V^{(0)}_{1/2^+}$ & $V^{(2)}_{1/2^+}$ & $E_{1/2^+}$ & ${\cal C}_{1/2^+}$ & ${\cal S}_{1s_{1/2} \otimes 0^+}$ \\ 
  & (MeV) & (MeV fm$^{-2}$) & (MeV) & (fm$^{-1/2}$) & \\ \hline
0 & $-7.771$ & $-27.192776$  & $-0.5031$ & 0.7861 & 0.80 \\
0.1 & $-14.879$ & $-23.57$ & $-0.5031$ & 0.7859  & 0.80 \\
0.3 & $-22.034$ & $-17.5$ & $-0.5031$ & 0.7860  & 0.77 \\
0.5 & $-15.941$ & $-15.2$ & $-0.5031$  & 0.7863 & 0.75 \\
0.7 & 21.286  & $-23.2$ & $-0.5031$  &0.7860 & 0.73\\ \hline
\textit{ab initio} & \cite{Calcietal16}  &   & $-0.5$ & 0.786 & 0.90  \\ 
\end{tabular}
\end{ruledtabular}
\end{table}

The effect of the degeneracy is, here too, seen in the eigenphaseshifts shown in Fig.~\ref{fig:gs_wf_down_15_22}(b).
With a small coupling ($\beta=0.1$, blue dashed line), we observe again a resonance-like behavior a bit higher than $E=1$~MeV.
The effect of that resonance is reduced for higher coupling.
Interestingly, the result at $\beta=0.5$ provides a phaseshift in fair agreement with the \emph{ab initio}'s.
However, this is at a cost of a radial wave function that matches the NCSMC prediction only for $r\gtrsim6$~fm.
Because this degeneracy has no physical rationale, and because this solution does not improve the effective description of the \emph{ab initio} overlap wave function, we do not deem it to be of any practical interest.

These results are independent of $\sigma$; similar results have been obtained with other values of the diagonal cutoff.
Accordingly, we can extend the conclusion of Sec.~\ref{type1sc} to this solution, and sum up that the solutions obtained with $\sigma_c>\sigma$ do not improve the agreement between our rotor-EFT and the \emph{ab initio} predictions.
This choice should not be considered in practical calculations.

\subsection{Bound excited state: $1/2^-$}\label{essc}

\begin{figure*}[ht] 
    \centering
    \begin{minipage}{0.50\linewidth}
        \centering
        \includegraphics[width=\linewidth]
        {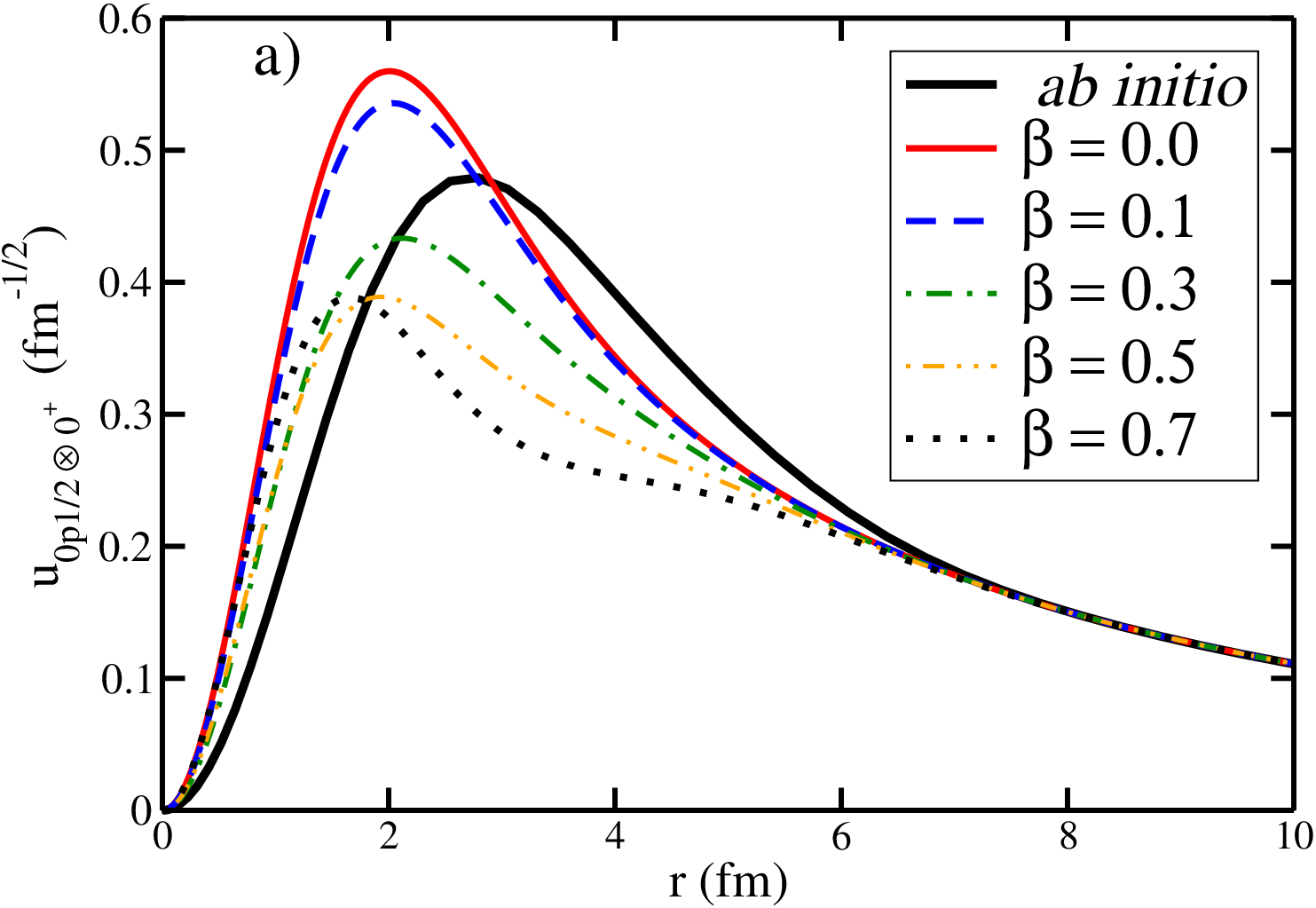}
    \end{minipage}\hfill
    \begin{minipage}{0.50\linewidth}
        \centering
        \includegraphics[width=\linewidth]
        {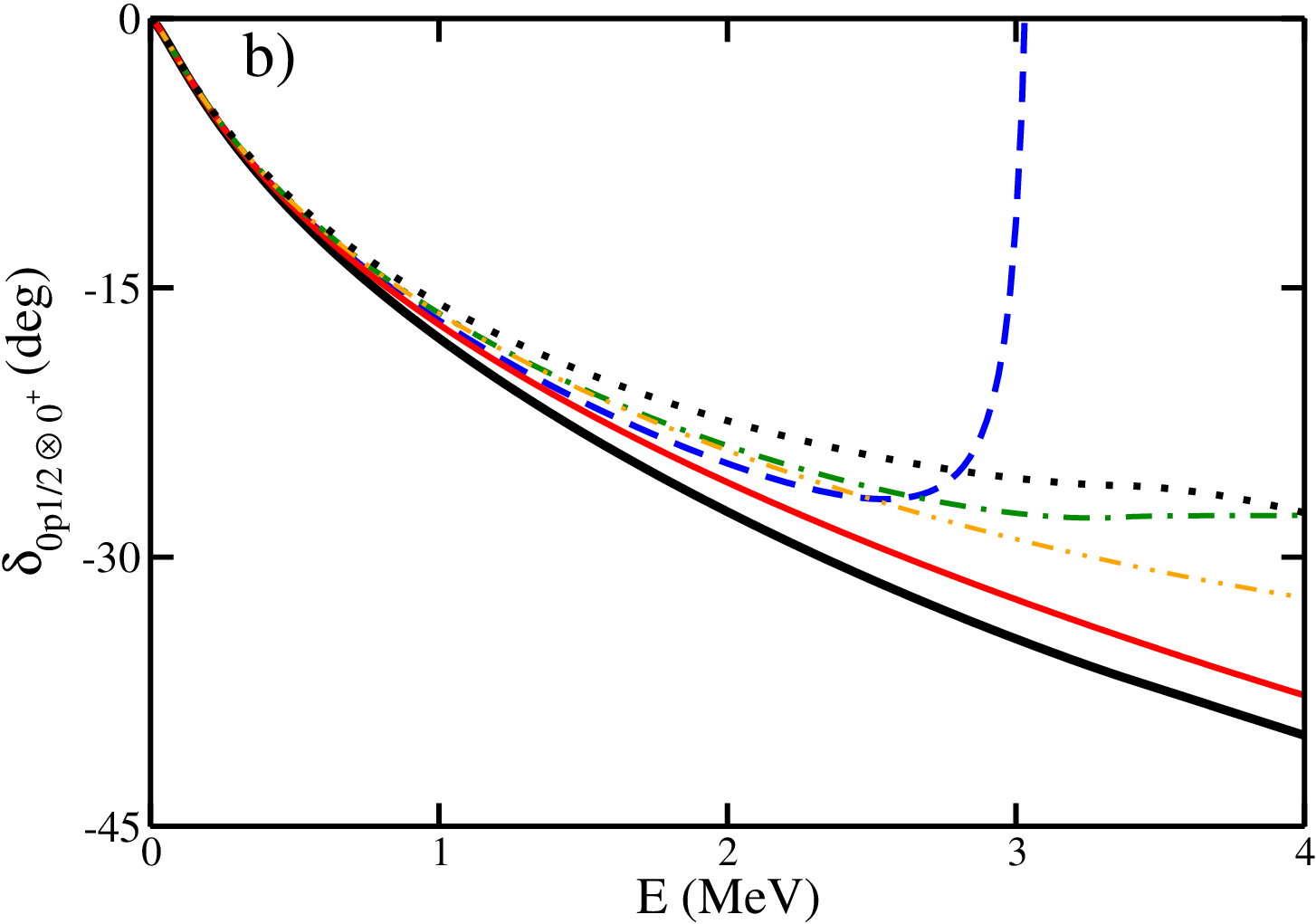}
    \end{minipage}
   \caption{Description of $^{11}$Be $\frac{1}{2}^-$ bound excited state in the extended rotor-EFT model of \Eq{eB1} with a coupling at $\sigma_c=2.2$~fm and a range of the diagonal potential of $\sigma=1.5$~fm.
    (a) Radial wave functions in the dominant $0p_{1/2} \otimes 0^+$ channel, computed for different values of the deformation parameter $\beta$. (b) Corresponding eigenphaseshifts in the $p_{1/2} \otimes 0^+$ channel as a function of the $^{10}$Be-$n$ scattering energy $E$. The \textit{ab initio} predictions of Ref.~\cite{Calcietal16} are also plotted in thick black lines.}
    \label{fig:es_wf_up_15_22}
\end{figure*}

The primary reason for using $\sigma_c>\sigma$ is to try to reproduce the overlap wave function of the $\frac{1}{2}^-$ excited bound state computed \emph{ab initio} within its \emph{pre-asymptotic} region.
The rationale behind that test is to generate additional attraction at the surface of the nucleus to increase the magnitude of the rotor-EFT radial wave function between 4 and 7~fm; see Fig.~\ref{fig:es_wf_up_15_15}(a).
Unfortunately, that trial fails, as shown in Fig.~\ref{fig:es_wf_up_15_22}(a) for $\sigma=1.5$~fm; similar results are obtained with other values of that cutoff.
The parameters of that new $c$-$n$ interaction are listed in Table~\ref{tab:es_wf_dep_1.5_2.2}.

\begin{table}[h]
\caption{\label{tab:es_wf_dep_1.5_2.2} 
Solution for the $\frac{1}{2}^-$ excited state obtained with the coupling shifted to larger radii. Parameters of the effective $^{10}$Be-$n$ potential \eqref{eB1} with $\sigma_c$=2.2~fm and $\sigma=1.5$~fm.
For each deformation parameter $\beta$, we provide the depths V$^{(0)}_{1/2^-}$ and V$^{(2)}_{1/2^-}$, which reproduce the listed $E_{1/2^-}$ energy, ANC, and SF.
The \textit{ab initio} predictions of Ref.~\cite{Calcietal16} are given in the last line.
}
\begin{ruledtabular}
\begin{tabular}{cccccccc}
 $\beta$ & $V^{(0)}_{1/2^-}$ & $V^{(2)}_{1/2^-}$ & $E_{1/2^-}$ & ${\cal C}_{1/2^-}$ & ${\cal S}_{0p_{1/2} \otimes 0^+}$ \\ 
  & (MeV) & (MeV fm$^{-2}$) & (MeV) & (fm$^{-1/2}$) & \\ \hline
0 & $-83.625$ & 5.2  & $-0.1841$ & 0.1291 & 1.00 \\
0.1 & $-82.419$ & 5.1 & $-0.1841$ & 0.1291  & 0.95  \\
0.3 & $-82.365$ & 6.7 & $-0.1841$ & 0.1291  & 0.73 \\
0.5 & $-98.044$ & 13.07 & $-0.1841$  & 0.1291 & 0.63 \\
0.7 & $-119.897$ &21.78 & $-0.1841$  &  0.1291 & 0.59 \\ \hline
\textit{ab initio} & \cite{Calcietal16}  &   & $-0.1848$  & 0.1291 & 0.85  \\ 
\end{tabular}
\end{ruledtabular}
\end{table}

As seen in \Sec{type1sc}, the behavior of the wave function is affected up to larger radii than in the case where $\sigma_c=\sigma$.
However, the peak of the radial wave function is not shifted to a larger radius, and the amplitude of the wave function between 4 and 7~fm is not increased.
To the contrary, because of the positive value of $V^{(2)}_{1/2^-}$, the maximum of our radial wave function is actually reached at lower $r$.
Moreover, as in the ground-state case, at large coupling parameter ($\beta=0.5$ and 0.7, dash-dot-dotted orange and dotted black lines, respectively), the wave function is distorted in the range $r\sim3$--4~fm.

The eigenphaseshifts shown in Fig.~\ref{fig:es_wf_up_15_22}(b) confirm that setting the coupling at the physical surface of the nucleus does not improve the Halo-EFT description of the \emph{ab initio} predictions.
From the results shown in Secs.~\ref{gssc} and \ref{essc}, we see that it is true for both bound states.
For the $\frac{1}{2}^-$ excited state in particular, it seems that, within the rotor-EFT presented here, no combination of parameters enables us to reproduce the \emph{ab initio} prediction of Calci \etal \cite{Calcietal16}.
For that state, it is better to simply use a N$^2$LO description of the nucleus as shown in Ref.~\cite{KC25}.

\providecommand{\noopsort}[1]{}\providecommand{\singleletter}[1]{#1}%

\end{document}